\documentclass[final,1p,times]{elsarticle}

\usepackage{graphicx}
\usepackage{amsmath}
\usepackage{amsbsy}
\usepackage{amssymb}
\usepackage{amscd}
\usepackage{amsfonts}
\usepackage{supertabular}
\usepackage{graphics}
\usepackage{graphicx}
\usepackage{subfigure}
\usepackage{epsfig}
\usepackage{float}
\usepackage{times}
\usepackage{epstopdf}
\usepackage{fancyhdr}
\usepackage{fancyheadings}
\usepackage{subfigure}
\usepackage{rotating}
\usepackage{pstcol}
\usepackage{pstricks}
\usepackage{color}
\usepackage{bm}

\newcommand{\mbs}[1]{\boldsymbol{#1}}

\def\bB{{\mbs{B}}}
\def\bF{{\mbs{F}}}
\def\bI{{\mbs{I}}}
\def\bK{{\mbs{K}}}
\def\bM{{\mbs{M}}}
\def\bN{{\mbs{N}}}
\def\bP{{\mbs{P}}}

\def\bS{{\mbs{S}}}
\def\bT{{\mbs{T}}}
\def\bQ{{\mbs{Q}}}
\def\bV{{\mbs{V}}}
\def\bX{{\mbs{X}}}

\def\be{{\mbs{e}}}

\def\bk{{\mbs{k}}}

\def\bu{{\mbs{u}}}
\def\bv{{\mbs{v}}}
\def\bx{{\mbs{x}}}

\def\b0{{\mbs{0}}}

\def\Fe {{\bF^{\rm e}}}
\def\Ff {{\bF^{\rm f}}}
\def\Fm {{\bF^{\rm m}}}
\def\D  {{\bm{\Delta}}}
\def\dt {{\Delta t}}
\def\la {{\bm{D}}}
\def\kf {{\bK^{\rm f}}}
\def\km {{\bK^{\rm m}}}

\def\vol{{L}}

\def\m {{^{\rm m}}}
\def\f {{^{\rm f}}}

\graphicspath {{./}{./Figures/}}

\graphicspath {{./}{./Figures/}}

\begin{document}

\begin{frontmatter}

\title{A multiscale model of distributed fracture and permeability in solids in all-round compression}

\author{Maria Laura De Bellis$^\star$}
\author{Gabriele Della Vecchia$^\dagger$}
\author{Michael Ortiz$^\ddagger$}
\author{Anna Pandolfi$^\dagger$ \corref{cor1}}
\ead{anna.pandolfi@polimi.it}
\cortext[cor1]{Corresponding author}
\address{
$^\star$ DII, Universit\`a~del Salento, 73100 Lecce, Italy\\
$^\dagger$ DICA, Politecnico di Milano, 20133 Milano, Italy\\
$^\ddagger$ EAS, Caltech, Pasadena CA 91125, USA}

\begin{abstract}
We present a microstructural model of permeability in fractured solids, where the fractures are described in terms of recursive families of parallel, equidistant cohesive faults. Faults originate upon the attainment of a tensile or shear resistance in the undamaged material. Secondary faults may form in a hierarchical organization, creating a complex network of connected fractures that modify the permeability of the solid. The undamaged solid may possess initial porosity and permeability. The particular geometry of the superposed micro-faults lends itself to an explicit analytical quantification of the porosity and permeability of the damaged material. The approach is particularly appealing as a means of modeling a wide scope of engineering problems, ranging from the prevention of water or gas outburst into underground mines to the prediction of the integrity of reservoirs for CO2 sequestration or hazardous waste storage.
\end{abstract}

\begin{keyword}
Microstructured permeability; parallel faults; multi-scale permeability; analytical models.
\end{keyword}

\end{frontmatter}

\graphicspath {{./}{./Figures/}}

\section{Introduction}

Damage induced by mechanical or hydraulic perturbations influences the permeability of the rock mass, with significant effects on the pore pressure distribution. Modifications in the pore pressure, in turn, affects the mechanical response of the material by poromechanical coupling. According to experimental observations at the microscopic scale, fracture evolution in rocks can be interpreted essentially as a progressive damage accumulation process, characterized by nucleation, growth and coalescence of numerous cracks following changes in the external load or in the internal pore pressure \cite{Wong_1996, Kranz_1983}. In the particular case of hydraulic fracturing, a stimulation technique used in petroleum industry to increase the oil/gas production in low permeability reservoirs, fractures are produced by the artificial increase of the fluid pressure in a borehole. From the theoretical point of view, it has been observed that the success of hydraulic fracturing is related to: i) the creation of a dense system of hydraulic cracks with limited spacing; and ii) the prevention or mitigation of localization instabilities \cite{Bazant_2014}.

Models of distributed damage and permeability based on abstract damage mechanics are, of necessity, empirical in nature and the precise meaning and geometry of the damage variables often remains undefined or is associated with unrealistic microstructures such as distributions of isolated microcracks. In addition, the evolution of the damage variables and their relation to the deformation, stress and permeability of the rock mass is described by means of empirical and phenomenological laws that represent, at best, enlightened data fits. However, the permeability enhancement due to extensive fracturing of a rock mass depends sensitively on precise details of the topology, which needs to be {\sl connected}, and geometry of the crack set, including the orientation and spacing of the cracks. In addition, the coupled hydro-mechanical response of the rock, especially when complex loading conditions and histories are of concern, is much too complex to yield to empirical data fitting.

Based on these considerations, in this paper we endeavor to develop a model of distributed fracturing of rock masses, and the attendant permeability enhancement thereof, based on an {\sl explicit micromechanical construction} of connected patterns of cracks, or faults. The approach extends the multi-scale brittle damage material model introduced in \cite{pandolfi:2006}, which is limited to mechanical damage. In contrast to abstract damage mechanics, the fracture patterns that form the basis of the theory are {\sl explicit} and the rock mass undergoes deformations that are compatible and remain in static equilibrium down to the micromechanical level. The fracture patterns are not arbitrary: they are shown in \cite{pandolfi:2006} to be optimal as regards their ability to relieve deviatoric stresses, and the inception, orientation and spacing of the fractures derive rigorously from energetic considerations. Following inception, fractures can deform by frictional sliding or undergo opening. The extension of the theory presented here additionally accounts for fluid pressure by recourse to Terzaghi's effective stress principle. When the fluid pressure is sufficiently high, existing fractures can open, thereby contributing to the permeability of the rock mass. By virtue of the explicit and connected nature of the predicted fractures, the attendant permeability enhancement can be estimated using simple relations from standard lubrication theory \cite{snow:1965, snow:1969, parsons:1966}, resulting in a fully-coupled hydro-mechanical model.

The paper is organized as follows. We begin in Section~\ref{sec:HydroMechanics} with illustrating the hydromechanical framework, recalling the basic equations and the Terzaghi's effective stress principle. In Section~\ref{sec:BrittleDamage} we recall the main features of the dry material model developed in \cite{pandolfi:2006}, introducing a pressure dependent behavior at fault inception. In Section~\ref{sec:Permeability} we derive analytically the permeability associated to the presence of faults in the brittle damage material model. In Section~\ref{sec:Examples} we validate the material model by means of comparison with experimental results taken from the literature.

\section{Hydro-mechanical framework}
\label{sec:HydroMechanics}

In porous media saturated with freely moving fluids, deterioration of mechanical and hydraulic properties of rock masses and subsequent problems are closely related to changes in the stress state, formation of new cracks, and increase of permeability. In fully saturated rocks, fluid and solid phases are fully interconnected and the interaction between fluid and rock is characterized by coupled diffusion-deformation mechanisms that convey an apparent time-dependent character to the mechanical properties of the system.

The two governing equations of the coupled problem are the linear momentum balance and the continuity equation (mass conservation). The kinematic quantities that characterize this picture are the porous solid displacement $\bu$ and the rate of fluid volume per unit area $q$. Hydro-mechanical coupling arises from the influence of the mechanical variables (stress, strain and displacement) on the continuity equation, where the primary variable is the fluid pressure, and from the influence of the hydraulic variables (pore pressure and seepage velocity) on the equilibrium equations, where the primary variables are the displacements.

\subsection{Continuity equation}
\label{ssec:FluidFlow}

The energy of a fluid flowing in a porous medium is traditionally measured in terms of total hydraulic head $h$, that for slow flowing fluids reads
\begin{equation}\label{eq:hydraulicHead}
    h = \frac{p}{\rho_f g} + z \, ,
    \nonumber
\end{equation}
where $\rho_f$ is the fluid density and $g$ the gravitational acceleration. The pressure head $p/\rho_f g$ is the equivalent gauge pressure of a column of water at the base of a piezometer. The elevation head $z$ expresses the relative potential energy. The kinetic energy contribution, $|\bv|^2/2$, is disregarded, given the small velocity $\bv$ of the fluid.

Flow across packed porous media is generally characterized by laminar regime (Reynolds number Re $\le$ 1) and by a drop of the hydraulic head in the direction of the flow.
Analytical models of fluid flow in rocks use constitutive relations that link the average fluid velocity across the medium to the hydraulic head drop. As representative example of constitutive relation in material form, Darcy's law states that the rate relative to the solid skeleton of the discharge per unit area of porous media, $Q$, is proportional to the hydraulic head gradient ${\mbs{\nabla}}_X \, h $ and inversely proportional to the fluid viscosity $\mu$
\begin{equation}\label{Eq:Darcy}
     \bQ = - \bK \, \frac{\rho_f g}{\mu} \,  {\mbs{\nabla}}_X \, h \, ,
\end{equation}
where $\bK$ denotes the material permeability tensor. Permeability measures the ability for fluids (gas or liquid) to flow through a porous solid material; it is intrinsically related to the void topology and does not account for the properties of the fluid. 
In anisotropic media, permeability is a symmetric (consequence of the Onsager reciprocal relations) and positive definite (a fluid cannot flow against the pressure drop) second order tensor $\bK$\textcolor{red}{, see  }. 
Real eigenvalues of the permeability tensor are the principal permeabilities, and the corresponding eigenvectors indicate the principal directions of flow, i.e., the directions where flow is parallel to the pressure drop. Clearly, fractures modify the permeability tensor, introducing new preferential directions for fluid flow.

Although affected by many factors, in non--fractured materials permeability is primarily related to the rock porosity (or void fraction) $n$, expressing the ratio between the volume of the voids $V_V$ and the total volume $V$, which includes also the volume of solids $V_S$
\begin{equation}\label{eq:porosity}
    n = \frac{V_V}{V} \, .
\end{equation}
In finite kinematics, the porosity is naturally associated to the jacobian of the deformation gradient $J=V/V_0$. By denoting the porosity of the stress-free material with $n_0$, it holds
\begin{equation}\label{eq:JacobianPorosity}
    n
    =
    \frac{1}{J} \frac{V_V}{V_0}
    =
    1 - \frac{1}{J} \left( 1 - n_0 \right) \, ,
\end{equation}
(see Appendix A for details of the derivation, also cf.~\cite{Borja:1995}). Note that, for very low values of $n_0$  and $J<1$, Eq.~\eqref{eq:JacobianPorosity} may provide negative values for $n$, thus a zero lower-bound must be enforced in calculations.

The rate of fluid volume $\bQ$ is linked to the porosity through the continuity equation, which for partially saturated voids in material form reads
\begin{equation}\label{eq:ContinuityEquationFull}
    \frac{\partial \left( J \, n \, S_r \, \rho_f \right) }{\partial t} =
    - \, {\rm Div} \, ( \rho_f \, \bQ ) \, ,
    \nonumber
\end{equation}
where $S_r$ the degree of saturation (i.~e., the fraction of the fluid volume), ${\rm Div}$ the material divergence operator, and $\partial / \partial t$ the partial derivative with respect to time. Under the rather standard assumption of fully saturated voids and incompressible fluid, the continuity equation becomes \cite{Borja:1995}
\begin{equation}\label{eq:ContinuityEquation}
    \frac{\partial \left( J \, n \right) }{\partial t} = - \, {\rm Div} \, \bQ \,.
    \nonumber
\end{equation}

\subsection{Mechanics equations}
\label{ssec:FluidFlow}

In the absence of any occluded porosity, the solid grains forming the matrix generally undergo negligible volume changes. In keeping with standard assumptions in geomechanics, we consider the solid phase of the matrix incompressible, thus we regard the change of the volume of the matrix as a change of the volume of the voids of the matrix. This assumption is consistent with the adoption of the Terzaghi's theory, used here in lieu of the more sophisticated Biot theory, chosen for the sake of simplicity and to limit the number of parameters. Moreover, we consider fully saturated media.

In finite kinematics the deformation is measured by the deformation gradient $\bF = {\partial \bx}/{\partial \bX}$, where $\bx$ and $\bX$ denote the spatial and material coordinates, respectively. The stress measure work-conjugate to $\bF$ is the first Piola-Kirchhoff tensor $\bP$. The linear momentum balance reads
\begin{equation}\label{eq:LinearMomentumFK}
    {\rm Div} \bP + \bB = \b0 \, ,
    \nonumber
\end{equation}
where $\bB$ is the material body force vector. Given the material traction $\bT$, the material boundary condition becomes
\begin{equation}\label{eq:BoundaryFK}
    \bP \bN = \bT \, .
    \nonumber
\end{equation}
In keeping with Terzaghi's principle of effective stress we write
\begin{equation}\label{eq:EffectiveStressFK}
    \bP = \bP' + p J \, \bF^{-T} \, ,
    \nonumber
\end{equation}
where $J$ is the determinant of $\bF$. The effective stress $\bP'$ and the deformation gradient define the constitutive law
\begin{equation}\label{eq:ConstitutiveFK}
    \bP' = \bP'(\bF) \, .
    \nonumber
\end{equation}

\section{Dry brittle damage model}
\label{sec:BrittleDamage}

\begin{figure}[!ht]
\begin{center}
    \vskip 0.3cm
    \subfigure[Reference configuration]{\epsfig{file=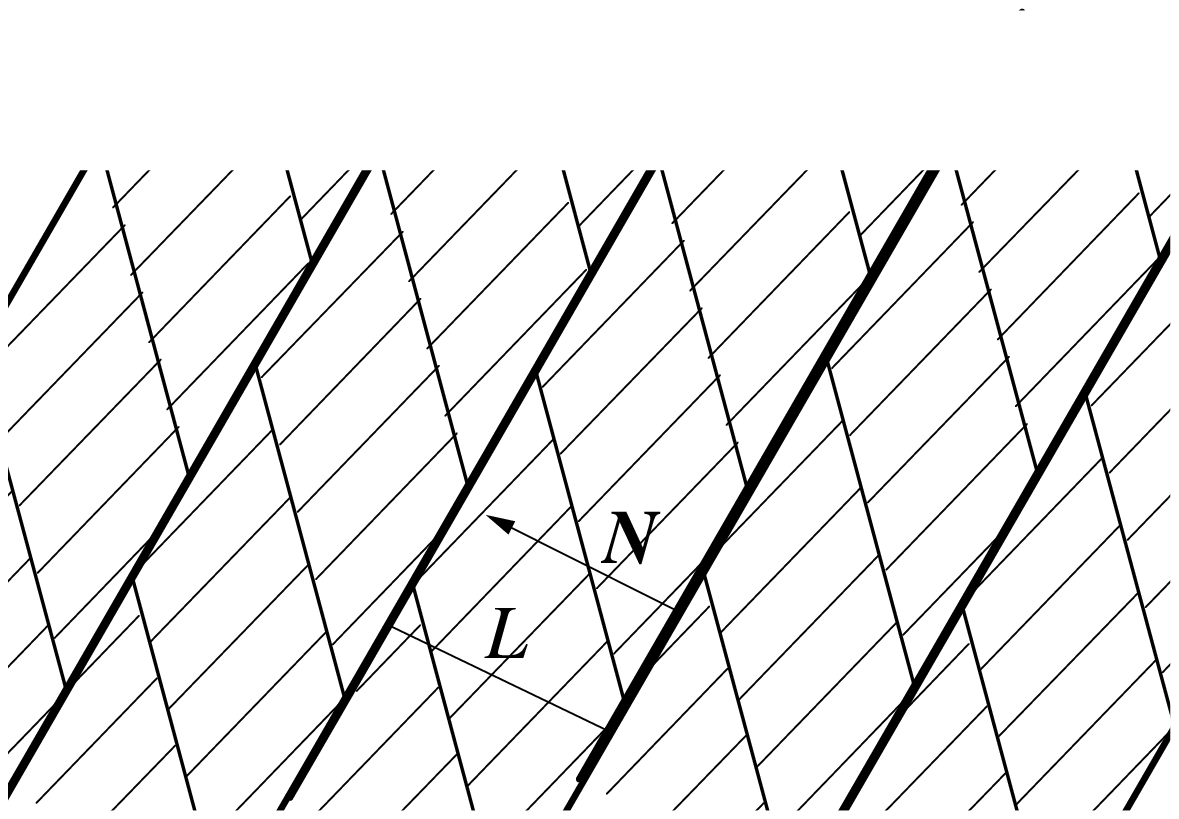,height=1.25in}\label{Fig:Fpa}}
    \hskip 0.15 \textwidth
    \subfigure[Current configuration]{\epsfig{file=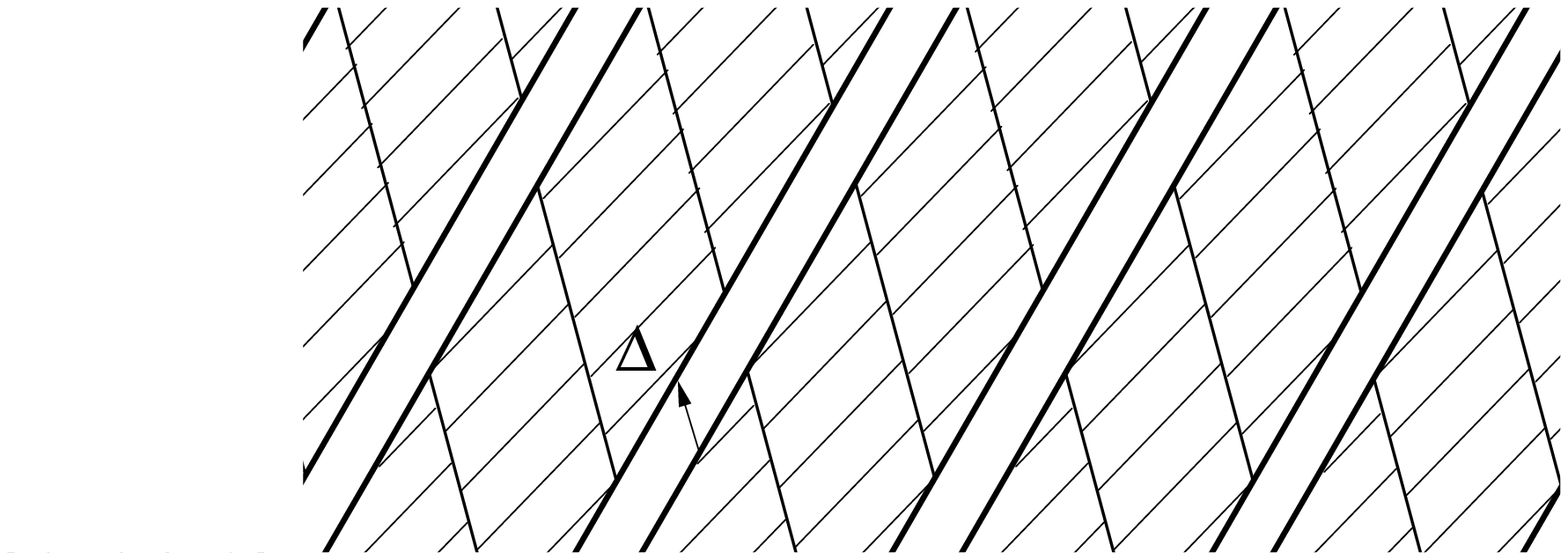,height=1.25in}\label{Fig:Fpb}}
    \caption[]{\small Inelastic kinematics of the rank-1 fault system. The opening displacement $\D$ applied to all the faults at distance $L$ leads to a deformed configuration characterized by the inelastic deformation gradient $\Ff$.}
    \label{Fig:Faults}
\end{center}
\end{figure}

The brittle damage model presented in \cite{pandolfi:2006} is characterized by a homogeneous matrix where nested microstructures of different length scales are embedded. At each level (or rank) $k$ of the nested architecture, microstructures assume the form of families of cohesive faults, characterized by an orientation $\bN_k$ and a uniform spacing $L_k$, see Fig.~\ref{Fig:Fpa}. In keeping with well established mathematical procedures used to treat free discontinuity problems, the brittle damage constitutive model is derived through a thermodynamically consistent approach, by assuming the existence of a free energy density which accounts for reversible and dissipative behaviors of the material.

\subsection{Recursive fault kinematics}
\label{ssec:kinematics}

The key of the brittle damage model is given by the kinematic assumptions. We begin by considering the particular case of a single family of fault planes of normal $\bN$ and spacing $L$, and later extend the behavior to recursive nested families.
The total deformation gradient $\bF$ of the material is assumed to decompose multiplicatively into a part $\Fm$ pertaining the uniform deformation of the matrix, and a part $\Ff$ describing the discontinuous kinematics of the cohesive faults, i.~e.,
\begin{equation}\label{eq:FmFf}
    \bF = \Fm \Ff \,.
    \nonumber
\end{equation}
The discontinuous deformation gradient $\Ff$ is related to the kinematic activity of the faults \cite{pandolfi:2006}, expressed through an opening displacement $\D$ acting on each fault of the family as (see Fig.~\ref{Fig:Fpb})
\begin{equation}\label{eq:Fc}
    \Ff
    \equiv
    \bI + \frac{1}{L} \, \D \otimes \bN \,.
    \nonumber
\end{equation}
Once $\bN$ and $L$ are supplied, $\Ff$ and $\D$ are in one-to-one correspondence.

The fractured material, in turn, may accommodate a second family of faults:
\begin{equation}\label{eq:FeFpLevels}
    \bF = {\Fm}^1 {\Ff}^1, \qquad {\Fm}^1 = {\Fm}^2 {\Ff}^2
    \nonumber
\end{equation}
This decomposition can be applied recursively for as many levels as necessary; the innermost level will maintain a purely elastic behavior $\Fe$.

\subsection{Constitutive assumptions}
\label{ssec:constitutiveAssumptions}

The constitutive behavior of the brittle damage model follows from the introduction of a free energy density sum of two contributions with full separation of variables
\begin{equation}\label{eq:FreeEnergy}
    A(\Fm, \D, q)
    =
    W\m(\Fm)
    +
    \frac{1}{L} \Phi(\D, q) \, ,
    \nonumber
\end{equation}
where $W\m$ is the strain-energy density per unit volume of the matrix, $\Phi$ is the cohesive energy per unit surface of faults, suitably divided by the length $L$ to provide a specific energy per unit of volume, $\D$ is the displacement jump, and $q$ is a scalar internal variable used to enforce irreversibility. Note that the separation of the variables excludes strong coupling between the two energies.

The operative form of the energy densities $W\m$ and $\Phi$ can be selected freely according to the particular material considered. In the present model, the cohesive energy of a fault with orientation $\bN$ is assumed to depend on an effective scalar opening displacement $\Delta$ defined as
\begin{equation}\label{Eq:EffectiveDelta}
    \Delta =
    \sqrt{(1 - \beta^2)\left(\D \cdot \bN\right)^2 + \beta^2 |\D|^2},
    \nonumber
\end{equation}
where $|\D|$ is the norm of the opening displacement and $\beta$ a material parameter measuring the ratio between the shear and tensile strengths of the material \cite{ortiz:1999b}. It follows that the cohesive behavior is expressed in terms of an effective cohesive law $\Phi(\D, q) = \Phi(\Delta, q)$, dependent on the effective opening displacement only. The effective traction $T$ is given by
\begin{equation}\label{Eq:TractionsScalar}
    T = \frac{\partial \Phi}{\partial \Delta}
    =
    \sqrt{
    \left(1 - \beta^{-2}\right) \left(\bT \cdot \bN \right)^2 + \beta^{-2} |\bT|^2
    }\, .
\end{equation}
\begin{figure}[!h]
\begin{center}
    \subfigure[]{\epsfig{file=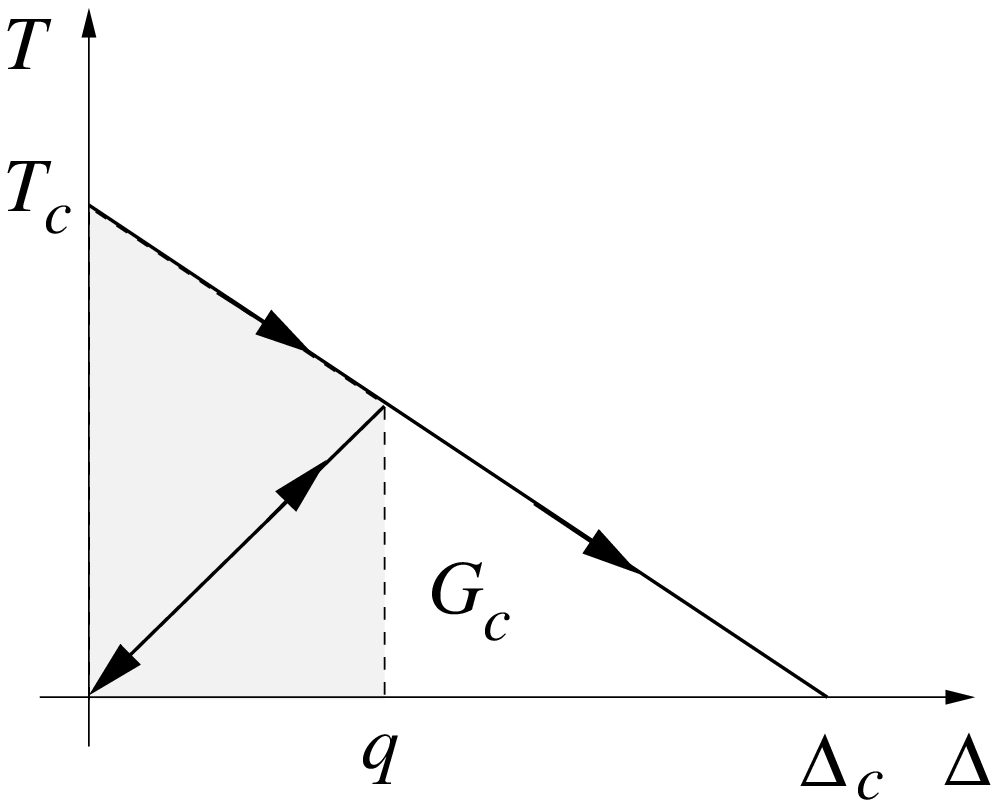,width=6cm}\label{Fig:LinearCohesive}}
    \hskip 0.5in
    \subfigure[]{\epsfig{file=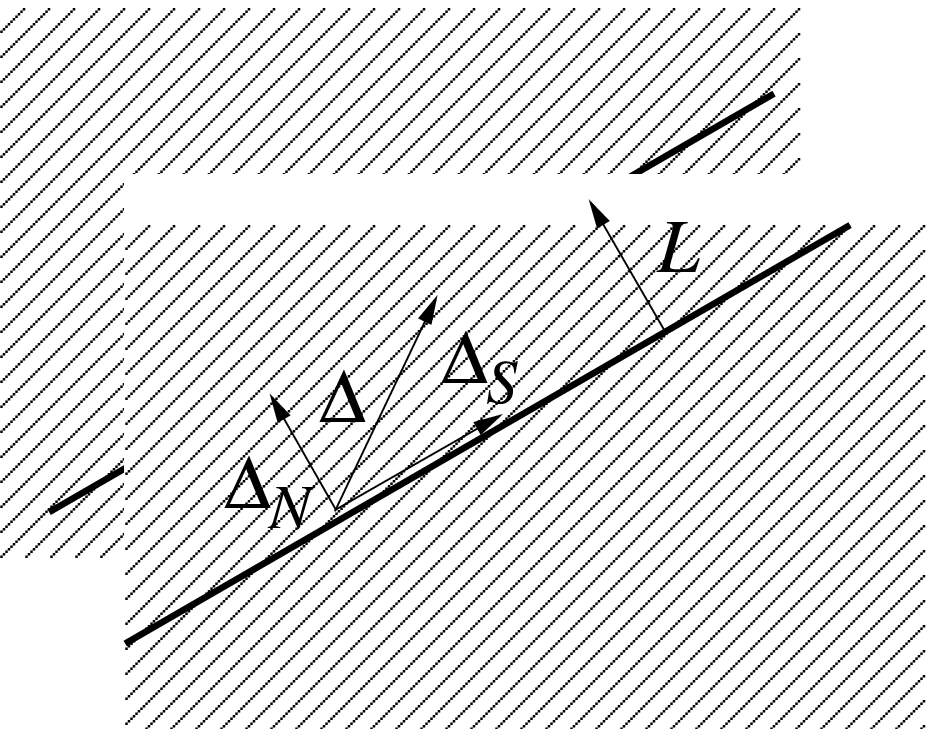,width=5cm}\label{Fig:jump}}
    \caption{\small (a) Irreversible linear decreasing cohesive law in terms of effective opening displacement and effective traction. The gray area represents the critical energy release rate $G_c$. The maximum traction is defined by the cohesive resistance $T_c$, and the maximum opening displacement is indicated by $\Delta_c$. The maximum attained effective opening $q$ defines the internal variable of the model, used to distinguish between first loading and unloading/reloading behaviors. (b) Kinematics of the single fault, defined by an opening displacement $\D$, with a component $\Delta_N$ along the normal and a component $\Delta_S$ in the plane of the fault.}
\end{center}
\end{figure}
In applications, we use a simple effective cohesive law, visualized Fig.~\ref{Fig:LinearCohesive}. During the first opening, the cohesive law follows a linearly decreasing envelope, i.~e.,
\begin{equation}\label{Eq:Cohesive}
    \Phi(\Delta, q) =
    T_c \Delta \left( 1 - 0.5 \, {\Delta}/{\Delta_c} \right)
\end{equation}
where $T_c$ the tensile resistance, $\Delta_c = 2 G_c / T_c$ the critical opening displacement corresponding to the full decohesion of the faults, and $G_c$ is the critical energy release rate of the material.
Tractions acting on the cohesive surface follows as, cf. \cite{pandolfi:2006},
\begin{equation}\label{Eq:Tractions}
    \bT
    =
    \frac{\partial \Phi}{\partial \D}
    =
    \frac{T}{\Delta}
    \left[
    \left(1 - \beta^2\right) \left(\D \cdot \bN \right) \, \bN + \beta^2 \D
    \right] \, .
\end{equation}
In the derivation of the constitutive model is necessary to introduce the configurational force conjugate to $\bN$, given by
\begin{equation}\label{Eq:Torquie}
    \frac{\partial \Phi}{\partial \bN}
    =
    \frac{T}{\Delta}
    \left(1 - \beta^2\right) \left(\D \cdot \bN \right) \, \D \, .
\end{equation}
Fracture is an irreversible process, thus decohered faults permanently damage the material. The extent of damage is expressed through the maximum attained effective opening displacement $q = \Delta_{\rm max}$. Irreversibility is enforced by assuming unloading and reloading to/from the origin, see Fig.~\ref{Fig:LinearCohesive}, according to the kinetic equation
\begin{equation}\label{Eq:kinetic}
    \dot q =
    \dot \Delta
    \quad {\rm if} \quad \Delta = q \quad {\rm and} \quad \dot \Delta \ge 0,
\end{equation}
Damage irreversibility is a constraint of the brittle damage model, enforced in calculations through the growth condition $\dot q > 0$. Upon fault closure
the material model has to satisfy the impenetrability constraint, i.~e., the component of the opening displacement along the normal to the faults cannot be negative, thus $\D \cdot \bN \ge 0$. More importantly, the model accounts for internal friction, a major dissipation mechanism in geological applications. We assume that friction operates at the faults concurrently with cohesion. Clearly, friction can become the sole dissipative mechanism if the faults loose cohesion completely upon the attainment of the critical opening displacement $\Delta_c$. In considering friction, we resort to the approach proposed in Pandolfi et al. \cite{pandolfi:2002} and make use of a dual dissipation potential per unit area $\Psi^*(\dot \D; \bF, \D, q)$, where $\dot \D$ denotes the rate of the fault opening displacement.

\subsection{Variational characterization}
\label{ssec:variationalCharacterization}

The behavior of irreversible materials with friction can be characterized variationally by recourse to time discretization \cite{ortiz:1999, pandolfi:2006}, where a process of deformation is analyzed at distinct successive times $t_0$, $\dots$, $t_{n+1} = t_n + \dt$, $\dots$. We assume that the state of the material at time $t_n$ ($\D_n$ and $q_n$) is known and the total deformation $\bF_{n+1}$ at time $t_{n+1}$ is assigned. The problem is to determine the state of the material at time $t_{n+1}$, accounting for material constraints and dissipation. We begin by considering a material with already a family of faults of spacing $L$ and orientation $\bN$.

Following \cite{ortiz:1999, pandolfi:2006}, the variational characterization of the material model requires to obtain an effective, incremental, strain-energy density $W_n(\bF_{n+1})$ by evaluating the infimum with respect to $\D_{n+1}$ and $q_{n+1}$ of the extended constrained energy defined as
\begin{equation}\label{eq:Wn}
    W_n(\bF_{n+1})
    =
    \inf_{
    \begin{matrix}
    \D_{n+1}, q_{n+1} \\ \D_{n+1} \cdot \bN \geq 0 \\ q_{n+1} \geq q_n
    \end{matrix}
    }
    A(\bF_{n+1}, \D_{n+1}, q_{n+1}) \,
    +
    \frac{\dt}{L} \,
    \psi^*\left(
    \frac{\D_{n+1} - \D_n}{\dt} ;
    \bF_{n+1}, \D_{n+1}, q_{n+1}
    \right).
\end{equation}
The subindex $n$ used in $W_n$ signifies the dependence on the initial state. The irreversibility and the impenetrability constraints render the effective strain-energy density $W_n$ dependent on the initial conditions at time $t_n$, and account for all the inelastic behaviors, such as damage, hysteresis, and path dependency. The constraints of the minimum problem can be enforced by means of two Lagrange multipliers $\lambda_1$ and $\lambda_2$, cf. \cite{pandolfi:2006}. Optimization leads to a system of four equations, that provide $\D_{n+1}$, $q_{n+1}$, $\lambda_1$, and $\lambda_2$. Thus, $W_n(\bF_{n+1})$ acts as a potential for the first Piola-Kirchhoff stress tensor $\bP_{n+1}$ at time $t_{n+1}$ \cite{ortiz:1999}, i.~e., as
\begin{equation}\label{eq:DWn}
    \bP_{n+1}
    =
    \frac{\partial W_n (\bF_{n+1})}
    {\partial\bF_{n+1}} \, .
\end{equation}
The stable equilibrium configurations are the minimizers of the corresponding effective energy.  Note that the variational formulation Eq.~\eqref{eq:Wn} of fault friction is non-standard in that it results in an incremental minimization problem. In particular, the tangent stiffness corresponding to the incremental equilibrium problem is symmetric, contrary to what is generally expected of non-associative materials. In calculations we assume rate independent Coulomb friction and, for the linearly decreasing cohesive model, we set
\begin{equation}\label{eq:Friction:Coulomb:EOS}
    \psi^*(\dot{\D}; \bS\m, \bN)
    =
    \mu_f \,\,
    {\rm max} \left\{
    0, \
     -
    \bN \cdot \bS\m \bN
    \right\}
    \,
    | \dot{\D} |
\end{equation}
where $\mu_f$ is the coefficient of friction and we denote the symmetric second Piola-Kirchhoff stress tensor of the matrix of components
\begin{equation}
    S^{\rm m}_{IJ}
    =
    F^{\rm m}_{iI} \frac{\partial W\m} {\partial F^{\rm m}_{iJ}} \,.
\end{equation}
The  dual dissipation  potential in Eq.~\eqref{eq:Friction:Coulomb:EOS} is rate-independent, i.~e., is positively homogeneous of degree $1$ in $\dot{\D}$, and proportional to the contact pressure.

\subsection{Fault inception and orientation for pressure dependent materials}
\label{ssec:FalutInception}

The fault geometrical features $\bN$ and $L$, which are defined by the surrounding stress state, can be determined with the aid of the time-discretized variational formulation, as described in \cite{pandolfi:2006}. The solution presented in \cite{pandolfi:2006} was addressing pressure independent materials under an extensive stress state. Here we provide a new solution, specific for stress states characterized by overall compression and for pressure sensitive materials.

Suppose that the material is undamaged at time $t_n$ and that we are given the deformation $\bF_{n+1}$ at time $t_{n+1}$. We test two end states of the material, one with faults and another without faults, and choose the end state which results in the lowest incremental energy density $W_n(\bF_{n+1})$. The time-discretized variational formulation allows to ascertain whether the insertion of faults is energetically favorable, and the optimal orientation of the faults in the fractured material. The orientation of the faults $\bN$ and the remaining state variables are obtained variationally from an extended constrained minimum problem, i.~e.,
\begin{equation}\label{eq:WnN}
    W_n(\bF_{n+1})
    =
    \inf_{
    \begin{matrix}
    \D_{n+1}, q_{n+1}, \bN \\
    \D_{n+1} \cdot\bN \geq 0 \\
    q_{n+1} \geq q_n \\
    |\bN|^2 = 1
    \end{matrix}
    }
    A(\bF_{n+1}, \D_{n+1}, q_{n+1}, \bN) \,
    +
    \frac{\dt}{L} \,
    \Psi^*
    \left(
    \frac{|\D_{n+1} - \D_n|}{\dt};
    \bS\m,
    \bN
    \right)   \, .
\end{equation}
Constrained optimization leads to a set of six equations, whose solution provides the optimal orientation $\bN$, $q$, $\Delta$, and three Lagrangian multipliers. For stress states in overall extension faults undergo simply opening. Thus, the frictional dissipation is null, and the resulting normal aligns with the direction of the maximum principal value of $\bS\m$, \cite{pandolfi:2006}.

With reference to pressure dependent materials in overall compressive states, the two optimization equations involving the normal $\bN$ --where we drop the index ${n+1}$ for the sake of clarity-- become:
\begin{subequations}
\begin{align}
    &
    \frac{\partial}{\partial \Delta_I}
    [A + \frac{\dt}{L} \Psi^* + \lambda_1 \D\cdot\bN]
    =
    - \frac{N_J}{L + \D\cdot\bN} \, S^{\rm m}_{JI}
    +
    \frac{1}{L}
    \frac{\partial\Phi}{\partial \Delta_I}
    +
    \frac{\dt}{L} \frac{\partial \Psi^*}{\partial \Delta_I}
    +
    \lambda_1 N_I
    = 0 \label{eq:X:Update:KT1}
    \\
    &
    \frac{\partial}{\partial N_I}
    [A + \frac{\dt}{L} \Psi^* + \lambda_1 \D\cdot\bN + \lambda_3 |\bN|^2]
    = \nonumber
    \\
    &
    - \frac{\Delta_J}{L + \D\cdot\bN} \,
    S^{\rm m}_{JI}
    +
    \frac{1}{L}\frac{\partial \Phi}{\partial N_I}
    +
    \frac{\dt}{L} \frac{\partial \Psi^*}{\partial N_I}
    +
    \lambda_1 \Delta_I
    +
    2 \lambda_3 N_I
    = 0  \, .
    \label{eq:X:Update:KT3}
\end{align}
\end{subequations}
Under a compressive stress, incipient faults are necessarily closed, $\D_n = \b0$, and can deform only by sliding, i.~e., $\D \cdot \bN = 0$. We denote with $\bM = \D/|\D|$ the unit vector in the direction of $\D$. Thus, the dissipation potential can be written as
\begin{equation}\label{eq:Friction:Coulomb:EOS:Inception}
    \Psi^*
    \left(
    \frac{|\D - \D_n|}{\dt};
    \bS\m,
    \bN
    \right)    =
    -
    \mu_f \,
    \bN \cdot \bS\m \bN
    \,
    \frac{| \D |}{\dt} \, ,
\end{equation}
$\D_n$ being null at the inception, and Eqs.~\eqref{eq:X:Update:KT1}-\eqref{eq:X:Update:KT3} become
\begin{equation}\label{eq:FrictionOptimal1}
    - S^{\rm m}_{IJ} N_J
    +
    \beta \, T
    M_I
    -
    \mu_f \, \bN \cdot \bS\m \bN \, M_I
    +
    L \, \lambda_1 N_I
    = 0
    \nonumber
\end{equation}
\begin{equation}\label{eq:FrictionOptimal2}
    -
    S^{\rm m}_{IJ} M_J
    -
    2 \, \mu_f \,  S^{\rm m}_{JI} N_J
    + L \, \lambda_1 M_I
    +
    \frac{2 \, \lambda_3 \, L}{|\D|} N_I
    = 0 \, .
    \nonumber
\end{equation}
Multiplying the first of these equations by $M_I$ and the second by $N_I$ we obtain the identities
\begin{equation}\label{eq:FrictionOptimal1-2}
    S^{\rm m}_{IJ} N_J M_I  + \mu_f \, S^{\rm m}_{IJ} N_J N_I
    =
    \beta T
    \nonumber
\end{equation}
\begin{equation}\label{eq:FrictionOptimal2-2}
    S^{\rm m}_{IJ} N_J M_I  + 2 \mu_f \, S^{\rm m}_{IJ} N_J N_I
    =
    \frac{2 \, \lambda_3 \, L}{|\D|} \, .
\end{equation}
The resulting equations imply that $\bN$ is a plane where the matrix shear stress satisfies the Mohr-Coulomb failure criterion, in the classical form
\begin{equation}\label{eq:Coulomb}
    \tau  = \beta T - \mu_f \, \sigma,
    \qquad
    \tau = S^{\rm m}_{IJ} N_J M_I,
    \qquad
    \sigma = S^{\rm m}_{IJ} N_J N_I \, ,
\end{equation}
where $T$ must be intended equal to $T_c$ at fault inception. Thus, when faults form, $\beta T_c$ corresponds to the cohesion (shear resistance at null normal stress) and $\mu_f = \tan \varphi$ the friction coefficient of the material. Eq.~\eqref{eq:Coulomb} sheds light on the meaning of the parameter $\beta$ that, for pressure sensitive materials, identifies with the friction coefficient $\mu_f$.
Finally, Eq.~\eqref{eq:FrictionOptimal2-2} provides the lagrangian multiplier $\lambda_3$ as
\begin{equation}\label{eq:FrictionOptimal2-3}
    \lambda_3
    =
    \frac{ |\D| }{2 \, L} \, \mu_f \, \left( T + \sigma \right) \, .
    \nonumber
\end{equation}
\\
Likewise, the length $L$ can be computed variationally by accounting for the misfit energy $E^{\rm mis}(\D, L)$ contained in the boundary layers that form at the junctions between faults and a confining boundary. In the model, the compatibility between the faults and their container is satisfied only on average, and this gives rise to boundary layers that penetrate into the faulted region to a certain depth. The addition to the energy furnishes a selection mechanism among all possible microstructures leading to a relaxed energy, cf. \cite{pandolfi:2006}.

So far we have been considering either an intact material or a single family of parallel faults. The material with a single fault family is referred to as rank-1 faulting pattern material. More complex microstructures can be generated effectively by applying the previous construction recursively.  In the first level of recursion, we simply replace the elastic strain-energy density $W\m(\Fm)$ of the matrix by $W_n(\Fm)$, i.~e., by the effective strain-energy density of a rank-1 faulting pattern. This substitution can now be iterated, resulting in a recursive definition of $W_n(\bF_{n+1})$. The recursion stops when the matrix between the faults remains elastic. The level of recursion is the rank of the microstructure. The resulting microstructures consist of faults within faults and are shown in Fig.~\ref{Fig:Faults}(a). Note that the implementation of the model in a numerical code is straightforward, and can be easily obtained by using recursive calls.

According to the particular loading history, at the time $t_n$ and at the generic point the material is be characterized by a particular microstructure with several $\D$, determined in respect of equilibrium and compatibility conditions. The model is therefore able to account for variable opening of the faults.

\section{Permeability of the brittle damage model}
\label{sec:Permeability}

Permeability is an overall important physical property of porous media very difficult to characterize theoretically. For simple and structured models of porous media, permeability can be estimated through analytical relationships that apply only under a narrow range of conditions.
The class of Kozeny-Carman type models collects simple relations that, under the assumption of laminar flow of the pore fluid, link the permeability to the microstructural characteristics of the porous medium. The original Kozeny-Carman relation \cite{Kozeny_1927, Carman_1937, Carman_1956} reads
\begin{equation}\label{Eq:KozenyCarman}
    k = \frac{c}{8 a_v^2 \tau} \, n \,  \left( \frac{n}{1-n}\right)^2
\end{equation}
where $k$ is a scalar permeability, $c$ an empirical geometric parameter, $a_v$ the ratio of the exposed surface of the channels to the volume of the solids (also called specific internal surface area), and $\tau$ the tortuosity, related to the ratio between $L_a$, average length of the channels, and $L$, macroscopic length of the flow path. The estimation of the shape coefficients $a_v$ and $\tau$  has been promoting an active research \cite{Goktepe_2010, Sanzeni_2013, Schaap_2001, Cieslicki_2000, Saripalli_2002}. The complexity of the relationship between the permeability tensor and a scalar property such as the porosity in rocks has been clearly pointed out \cite{Schulze_2001}.

The scalar nature of variables and parameters used in analytical models leads to scalar definitions, and the correct tensor nature of the permeability is disregarded. Therefore, such models are not meaningful if applied to soils characterized by the presence of sedimentation layers or fissures. Moreover, these models do not allow for the modification of the porous medium microstructure due to fluid-porous matrix interactions, or by the presence of a variable confining pressure. In particular, permeability depends not only on the actual stress and on the strain during the loading history, but also on the evolution of the crack patterns, which is anisotropic in nature.

Considering the presence of a single fault family, the permeability tensor for the fractured brittle damage model due to the sole presence of the faults can be directly derived from the particular faults geometry. The permeability of a particular geometry of parallel and equidistant faults has been examined by Irmay \cite{irmay:1955}. Snow \cite{snow:1965,snow:1969} and Parsons \cite{parsons:1966} obtained expression for anisotropic permeability, similar to the one described here, by considering networks of parallel fissures.

We begin by recalling that the opening displacement decomposes into a normal $\Delta_N$ and a sliding $\Delta_S$ components, see Fig.~\ref{Fig:jump}, computed as:
\begin{equation}\label{eq:DeltaN}
    \Delta_N
    =
    \bN \cdot \D,
    \qquad
    \D_S
    =
    \left( \bI - \bN \otimes \bN \right) \,
    \D,
    \qquad
    \Delta_S = |\D_S|,
\end{equation}
Let us assume that a fluid flows within the faults, filling the open layers of constant width $\Delta_N$. The average fluid flow, in laminar regime, will take place in the plane of the layer. According to the solution of the Navier-Stokes' equation (Poiseuille's solution), the average velocity $V_s$ along the generic direction $s$ in the plane of the fault is
\begin{equation}\label{eq:fissureVelocity}
    V_s
    =
    -
    \frac{\Delta_N^2}{12} \, \frac{\rho_f g}{\mu} \frac{\partial h}{\partial s} \, ,
\end{equation}
where ${\partial h}/{\partial s}$ is the hydraulic head gradient in the direction $s$. The assumption of laminar flow through a crack has been widely used in the literature, cf., e.~g., \cite{Peirce_2008}. By considering a porous medium made of several parallel faults of equal width, the discharge $Q_s$ in the direction of the flow is
\begin{equation}\label{eq:fissureDischarge}
    Q_s
    =
    n\f \, V_s
    =
    -
    \frac{\Delta_N}{\vol} \, \frac{\Delta_N^2}{12}\, \frac{\rho_f g}{\mu}
    \frac{\partial h}{\partial s},
\end{equation}
where
\begin{equation}\label{eq:fissurePorosity}
    n\f = \frac{\Delta_N}{\vol}
\end{equation}
is a measure of the material porosity due exclusively to the presence of faults. By comparing Eqs.~\eqref{eq:fissureDischarge} and \eqref{Eq:Darcy}, we obtain the permeability $K_s$ of the fractured material in direction $s$ as
\begin{equation}\label{eq:permeabilityBD}
    K_s
    =
    \frac{\Delta_N^3}{12 \, \vol} .
\end{equation}
The directional gradient $\partial h/\partial s$ can be expressed as the scalar product of the hydraulic gradient ${\bm\nabla}_X\, h$ and the flow direction $\la$, so that the magnitude of the fluid velocity reads
\begin{equation}\label{eq:fissureVelocityScalar}
    V_s
    =
    -
    \frac{\Delta_N^2}{12}\, \frac{\rho_f g}{\mu} \, {\mbs{\nabla}}_X \, h \cdot \la \, ,
\end{equation}
and the average flow velocity vector, $\bV_s = V_s \la$, becomes
\begin{equation}\label{eq:fissureVelocityVector}
    \bV_s
    =
    -
    \frac{\Delta_N^2}{12}\, \frac{\rho_f g}{\mu} \left( {\mbs{\nabla}}_X \, h \cdot \la \right) \, \la.
\end{equation}
The hydraulic discharge can be written as
\begin{equation}\label{eq:fissureDischargeGlobal}
    \bQ_s
    =
    n\f \, \bV_s
    =
    -
    \frac{\Delta_N^3}{12 \, \vol}
    \left(
    \la
    \otimes
    \la
    \right)
    \,
    \frac{\rho_f g}{\mu}
    \,
    {\mbs{\nabla}}_X \, h \, ,
\end{equation}
thus the permeability tensor due to the presence of the faults $\bk^{\rm f}$ derives as
\begin{equation}\label{eq:permeabilityTensor1}
    \bK^{\rm f}
    =
    \frac{\Delta_N^3}{12 \, \vol}
    \,
    \left(
    \la
    \otimes
    \la
    \right).
\end{equation}
To account for a generic direction of the flow in the layer of normal $\bN$, in Eq.~\eqref{eq:permeabilityTensor1} we must replace the unit vector $\la$ with  the projection $(\bI - \bN \otimes \bN)$, reaching the expression
\begin{equation}\label{eq:permeabilityTensor2}
    \kf
    =
    \frac{\Delta_N^3}{12 \, \vol}
    \,
    \left(
    \bI
    -
    \bN
    \otimes
    \bN
    \right).
\end{equation}
As a noteworthy feature of the brittle damage model, it follows that the permeability is described by an anisotropic tensor.

If $Q$ fault families are present in the porous medium, each characterized by a normal $\bN^K$, a separation $L^K$, and a normal opening displacement $\Delta^K_N$, the equivalent permeability is given by the sum of the corresponding permeabilities:
\begin{equation}\label{eq:permeabilityTensorEquiv}
    \kf
    =
    \sum_{K=1}^Q {
    \frac{{\Delta_N^K}^3}{12 \, \vol^K}
    \,
    \left(
    \bI
    -
    \bN^K
    \otimes
    \bN^K
    \right)}.
\end{equation}
The model does not exclude the presence of an initial porosity $n\m$, see Eq.~\eqref{eq:porosity}, and permeability $\km$, see Eq.~\eqref{Eq:KozenyCarman}, of the intact matrix. In this case, the resulting porosity and permeability will be given by the sum of the terms corresponding to the intact matrix and to the faults
\begin{equation}\label{eq:permeabilityTensor3}
    n
    =
    n\m
    +
    n\f \, ,
    \qquad
    \bK
    =
    \km
    +
    \kf.
    \nonumber
\end{equation}
In practical applications we assume an isotropic matrix permeability of Kozeny-Carman type, with the simplified form
\begin{equation}\label{Eq:KozenyCarmanPermeability}
    \km = K_{\rm KC} \bI \, ,
    \qquad
    K_{\rm KC} = C_{\rm KC} \, \frac{({n\m})^3}{(1-n\m)^2} \, ,
\end{equation}
where the constant $C_{\rm KC}$ accounts for shape coefficients.

We observe that the hydraulic behavior of the brittle damage model is dependent on fracture orientation and spacing computed on the basis of the boundary conditions, and that its permeability can vary according to the kinematics of the faults.

We remark that the solid phase incompressibility assumption adopted in the present model is not affecting substantially the hydraulic behavior, mostly because the porosity of the matrix plays a minor role in the hydraulic conductivity of the material. In fact in this model the porosity, and thus the permeability, is mostly imputable to the formation of faults, downsizing the relevance of the matrix porosity.

\section{Verification and Validation}
\label{sec:Examples}

Numerical calculations of the dynamic multiaxial compression experiments on sintered aluminum nitride (AlN) of Chen and Ravichandran \cite{chen:1994, chen:1996, chen:1996b, chen:2000} were presented in \cite{pandolfi:2006} by way of validation of the dry mechanical aspects of the model. The model was shown to correctly predict the general trends regarding the experimental observed damage patterns, as well as the brittle-to-ductile transition resulting under increasing confinement. Therefore, in the present work we restrict validation to the hydro-mechanical aspects of the model. We describe selected examples of application of the porous damage model, starting from the response of the fully tridimensional dry model undergoing a loading that mimics a hydraulic fracturing process, and concluding with the validation of the model, reproducing a few representative experimental results on granite and sandstone.

\subsection{Illustrative dry example}
\label{ssec:dryExamples}
\begin{figure}[!ht]
\begin{center}
    \subfigure[Stress-Strain]
    {\epsfig{file=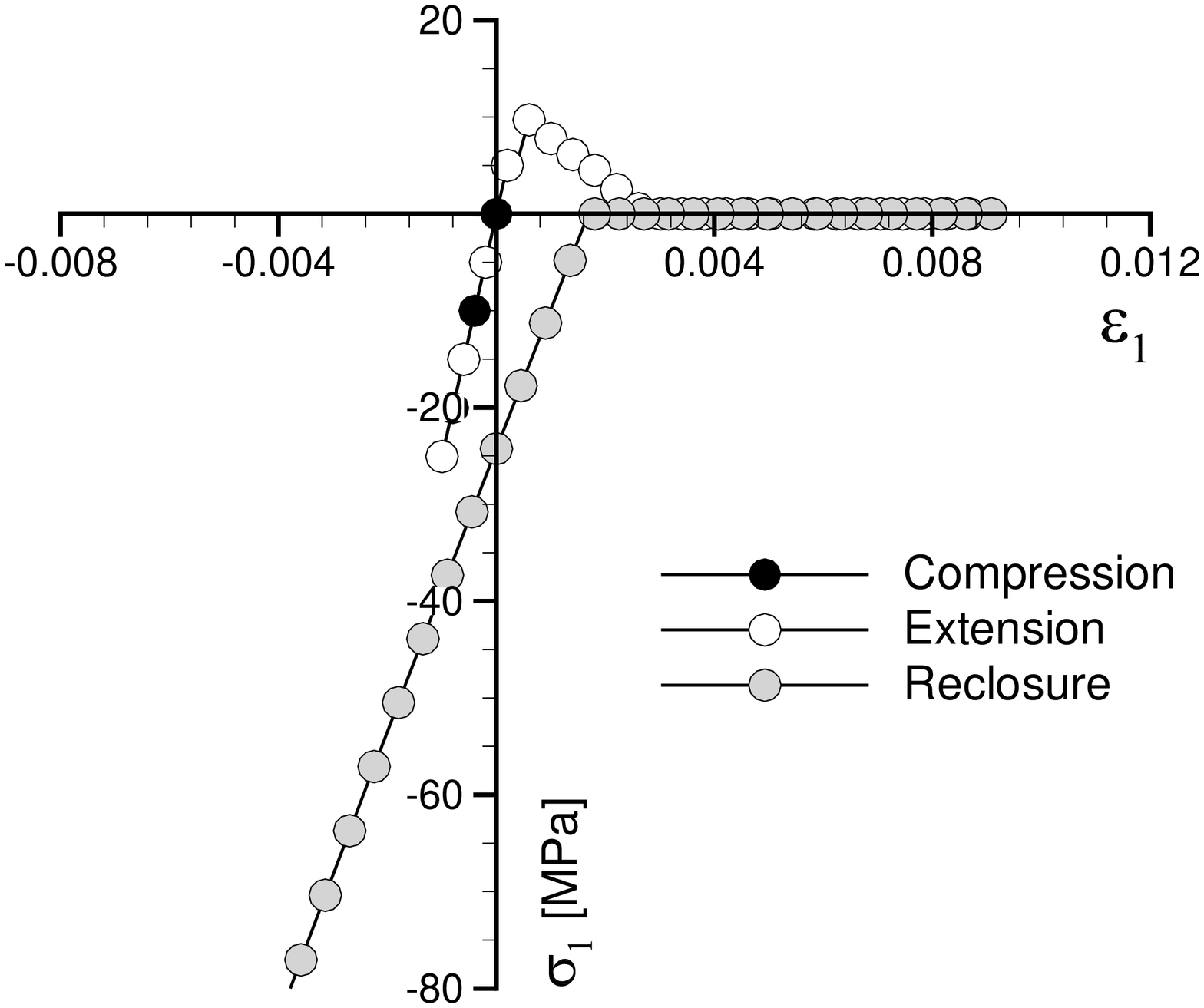,width=0.48\textwidth}\label{Fig:FrackingStress}}
    \subfigure[Permeability $k_{11}$] {\epsfig{file=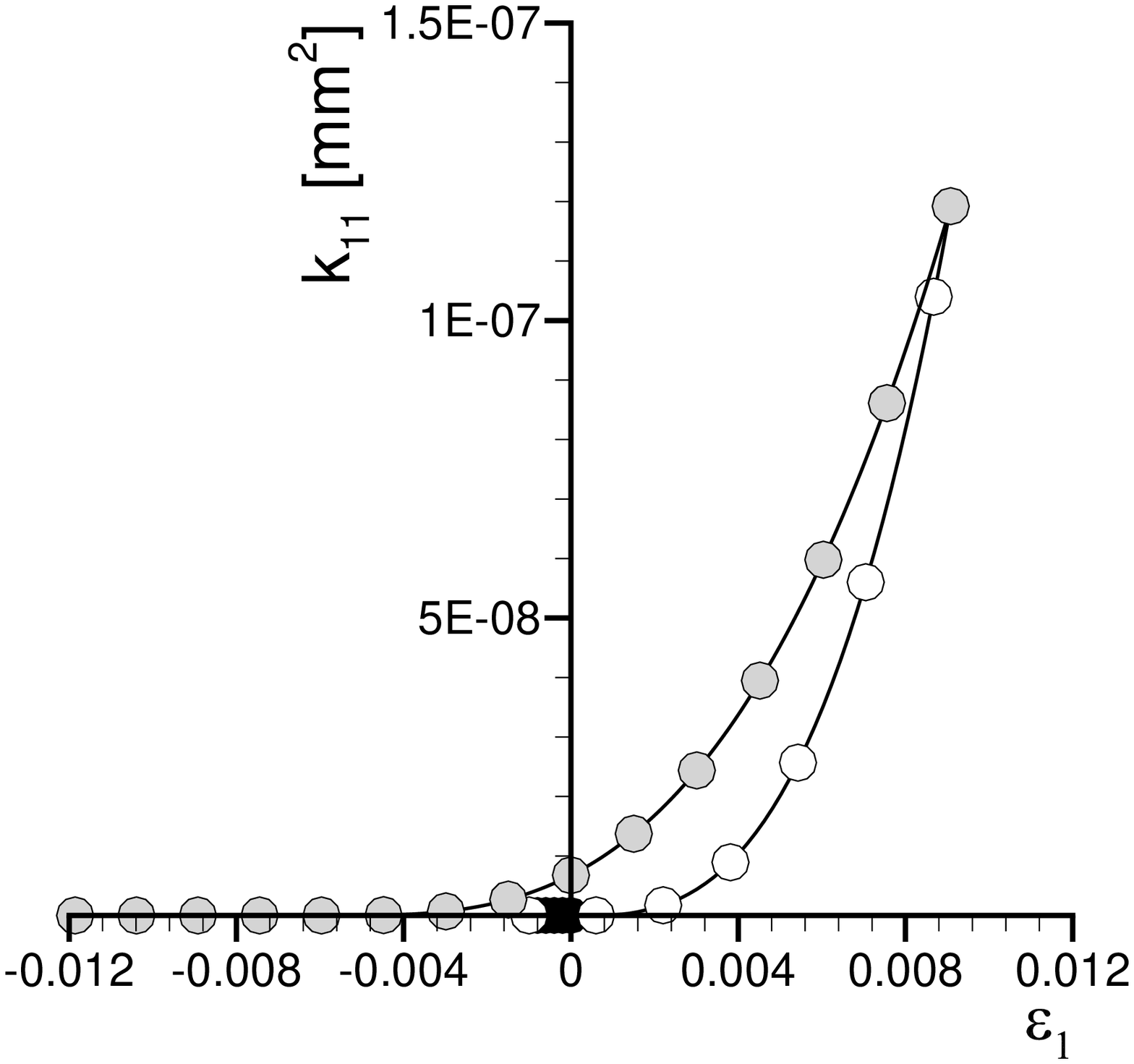,width=0.43\textwidth}\label{Fig:FrackingPerm1}}
    \subfigure[Permeability $k_{22}$]
    {\epsfig{file=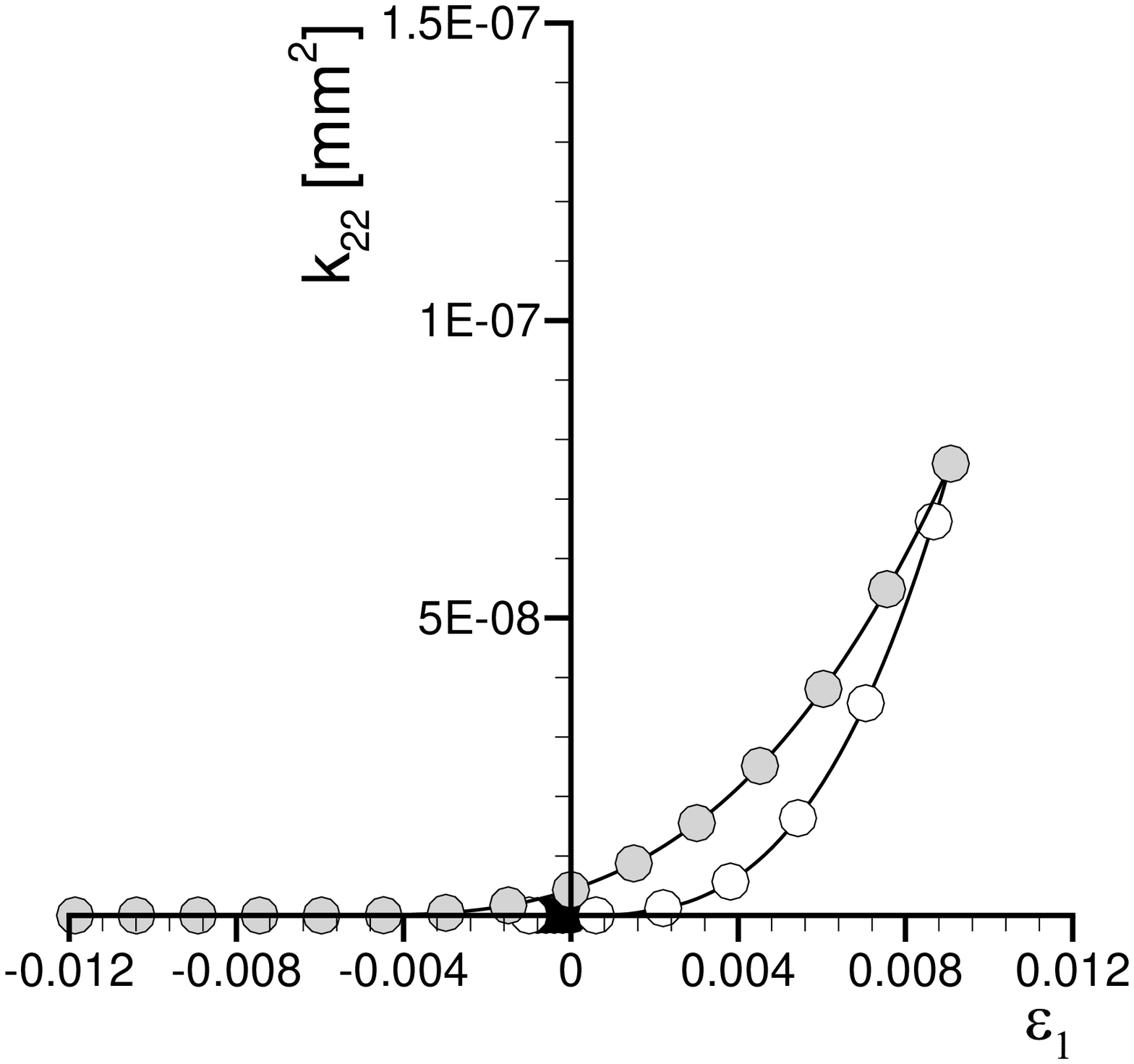,width=0.48\textwidth}\label{Fig:FrackingPerm2}}
    \subfigure[Permeability $k_{33}$]
    {\epsfig{file=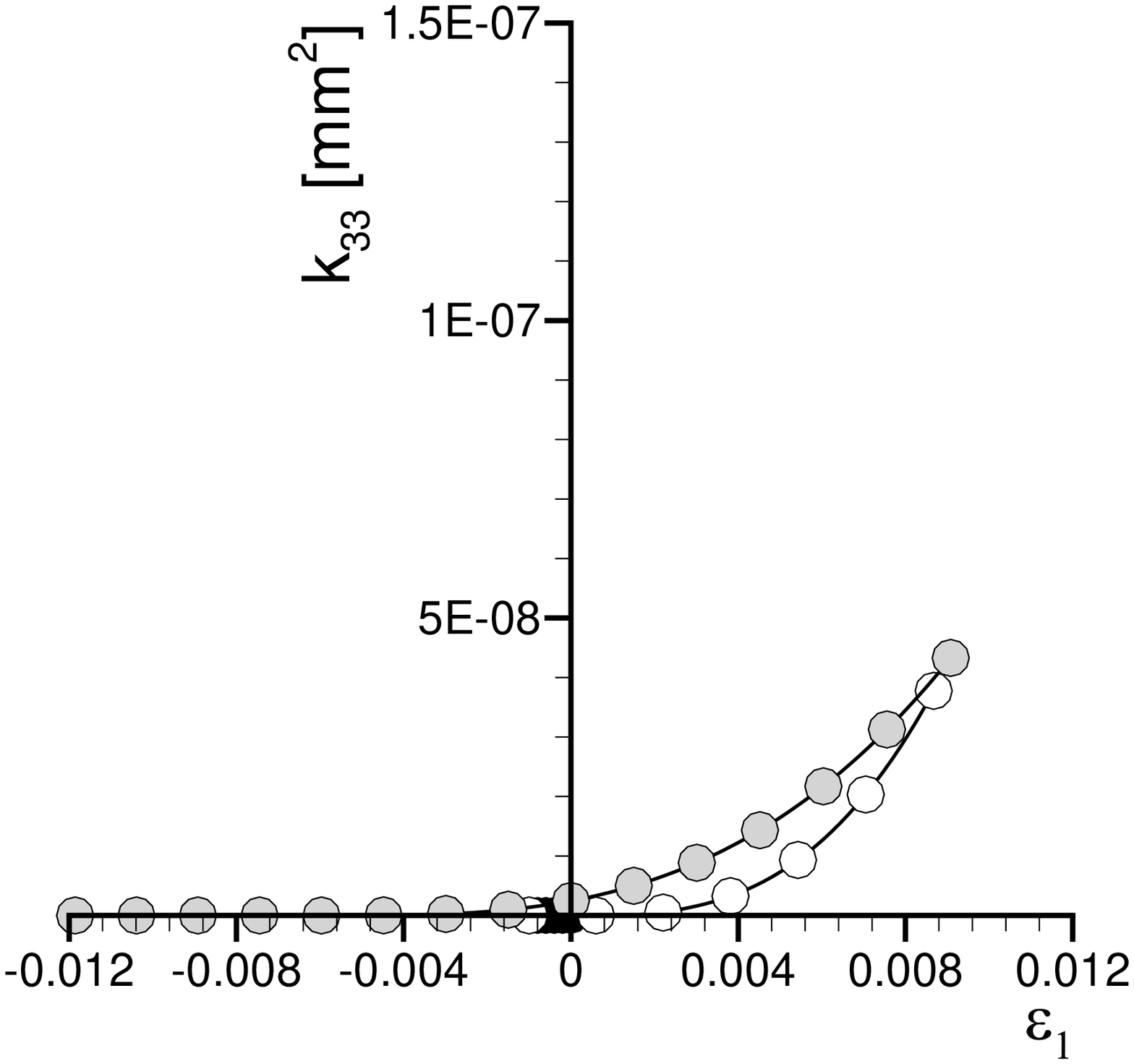,width=0.43\textwidth}\label{Fig:FrackingPerm3}}
    \caption{\small Multi-stage multiaxial response of the material, undergoing an isotropic compression, followed by an isotropic extension, and by a final anisotropic compression. (a) Stress $\sigma_1$ (MPa) vs strain $\varepsilon_1$. Although the material fails, generating three nested families of faults, the material preserves its ability of sustaining load.
    (b) Permeability (mm$^2$) in direction $\be_1$ as a function of the strain.
    (c) Permeability (mm$^2$) in direction $\be_2$ as a function of the strain.
    (d) Permeability (mm$^2$) in direction $\be_3$ as a function of the strain.}
    \label{Fig:CompressiveTest}
\end{center}
\end{figure}
We specialize the strain energy density $W$ to a neo-Hookean material extended to the compressible range, i.~e.,
\begin{equation}\label{Eq:NeoHookean}
    W(\Fm) =
    \frac{1}{2} \lambda \log^2{J\m}
    +
    \frac{1}{2} G \left( ({\Fm}^T \Fm) : \bI - 3 - 2 \log{J\m} \right)
\end{equation}
where $\lambda$ and $G$ are the Lam\'e coefficients, and $J\m = \det{\Fm}$ is the determinant of $\Fm$. We study the response of the brittle damage model to the action of external loadings mimicking the in-field conditions observed during hydraulic fracturing procedures, and analyze the correspondent variation in permeability. We assume an intact material, with no pre-existent or natural faults, and limit our attention to the constitutive behavior. The material is characterized by the constants listed in Table~\ref{Table:PropertiesFracking}.
\begin{table}[h]
\begin{center}
\caption {Rock material constants adopted in the illustrative examples}
\vskip 0.1in
\begin{tabular}{cccccccccccc}
    \hline
    $\lambda$ (MPa) & $G$ (MPa) & $T_c$ (MPa) & $G_c$ (N/mm) & $\varphi$ &  $k_0$ & $n_0$ & $C_{\rm KC}$ \\ \hline
    2778 & 4167 & 10.0 & 0.1 & 45.0 & 0 & 0 & 0 \\ \hline
\end{tabular}
\label{Table:PropertiesFracking}
\end{center}
\end{table}
The porosity and the permeability of the intact matrix are assumed to be null, thus the permeability will be exclusively related to the formations of faults. The material is allowed to form up to three families of faults, with different orientation $\bN^K$. Tensile stresses and deformations are considered positive.

By assigning a prescribed history to the deformation gradient, we simulate a multistage multiaxial test that mimics the in-field variation in stress and permeability due to hydraulic fracture. The material is initially compressed isotropically by applying a uniform stretch $L/L_0 = \lambda_1 = \lambda_2 = \lambda_3 = 0.99$, to induce a geostatic-like stress state. Then, the material undergoes an isotropic extension $\lambda_1 = \lambda_2 = \lambda_3 = 1.01$, associated to the reduction of the effective stress due to the injection of a high pressure fracturing fluid. Given the isotropy of the stress state, at the extension corresponding to the attainment material strength the material fails in tension, creating in sequence three families of faults, with normal in the directions $\be_1$, $\be_2$, and $\be_3$, respectively. The microstructure of the three families differs because of different spacings. Upon fault closure, the failed material is able to sustain an overall compressive stress, the interpenetration of the faults being controlled by the contact algorithm. In the last stage of loading, the material is compressed with an anisotropic stretch. A $\lambda_1 = \lambda_2 = 0.97$ stretch is applied in direction $\be_1$ and $\be_2$, while the original geostatic-like stretch $\lambda_3 = 0.99$ is applied in direction $\be_3$. Fig.~\ref{Fig:CompressiveTest}(a) shows the mechanical response of the model in direction $\be_1$. The material initially undergoes a compression (black circles). The following extension induces a tensile state that reaches the strength of the material and causes triple tensile failure (open circles). The final compressive stretch is characterized by a null stress until faults close completely. Afterwards, the contact algorithm provides the compressive tractions that guarantee the equilibrium of the system (grey circles). The resulting reduction of the stiffness of the material due to the damage is remarkable.

Figs.~\ref{Fig:CompressiveTest}(b-d) show the permeability in direction $\be_1$, $\be_2$, and $\be_3$, respectively. The permeability is null until the material fails (black circles). Then the permeability reaches a maximum corresponding to the maximum extension imposed to the material (open circles). The different values of the maximum permeability for the three directions is the combined result of the different spacing of the fault families and of the stress anisotropy derived from the formation of faults. Upon fault closure, permeability decreases to zero (gray circles). Note that the anisotropy of the compressive loading causes anisotropy in the permeability history. In particular, the permeability reduces more quickly in the direction $\be_3$, where no extra-confinement is applied. In fact, the over-compression in the two directions $\be_1$ and $\be_2$ closes the faults parallel to direction $\be_3$, while the flow is still allowed in the faults normal to $\be_3$.

\subsection{Validation of the porous model against experimental results}

\label{ssec:validation}
\begin{figure}[!ht]
\begin{center}
    \subfigure[Lac du Bonnet granite stress-strain]
    {\epsfig{file=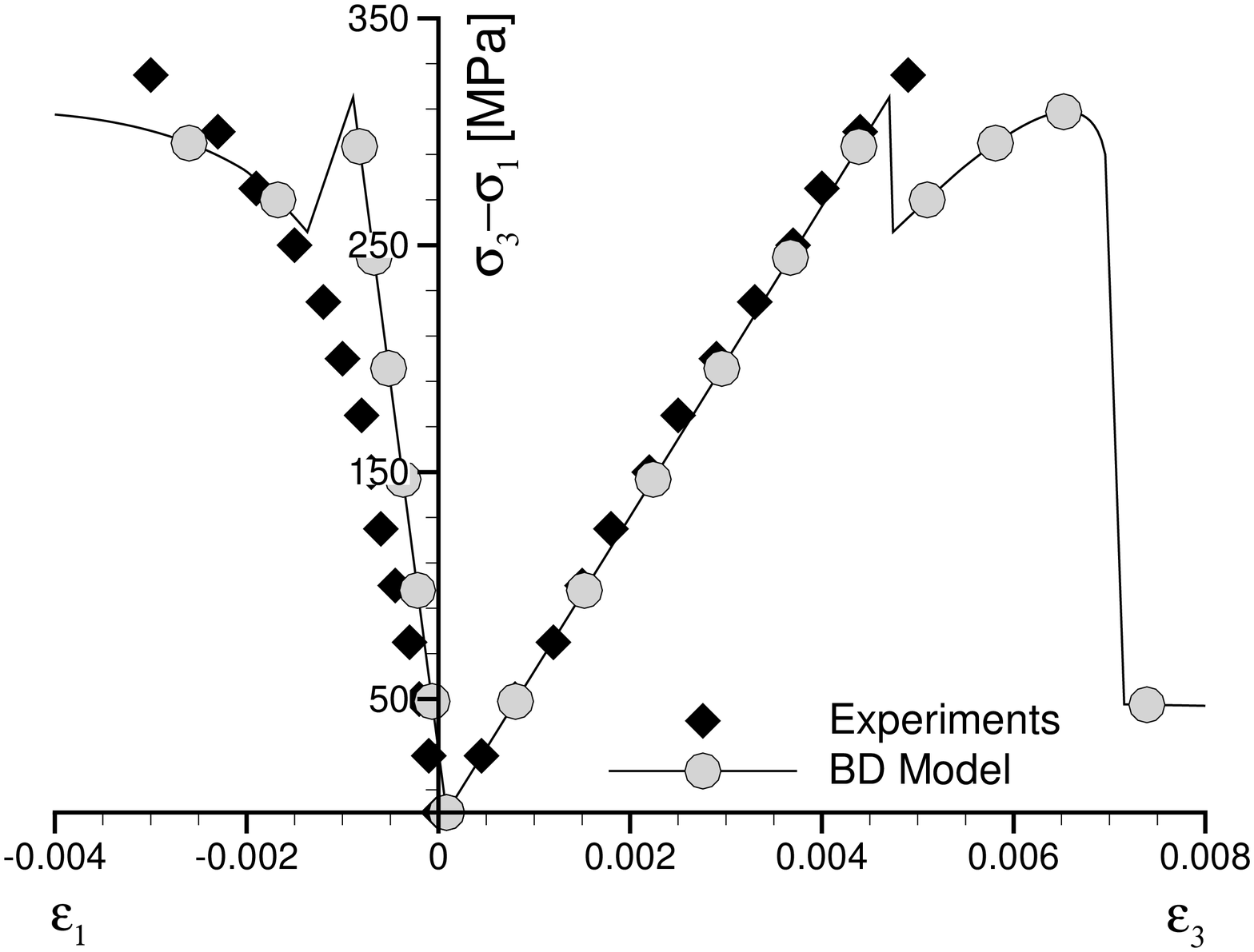,width=0.48\textwidth}\label{Fig:LacDuBonneta}}
    \subfigure[Lac du Bonnet granite permeability] {\epsfig{file=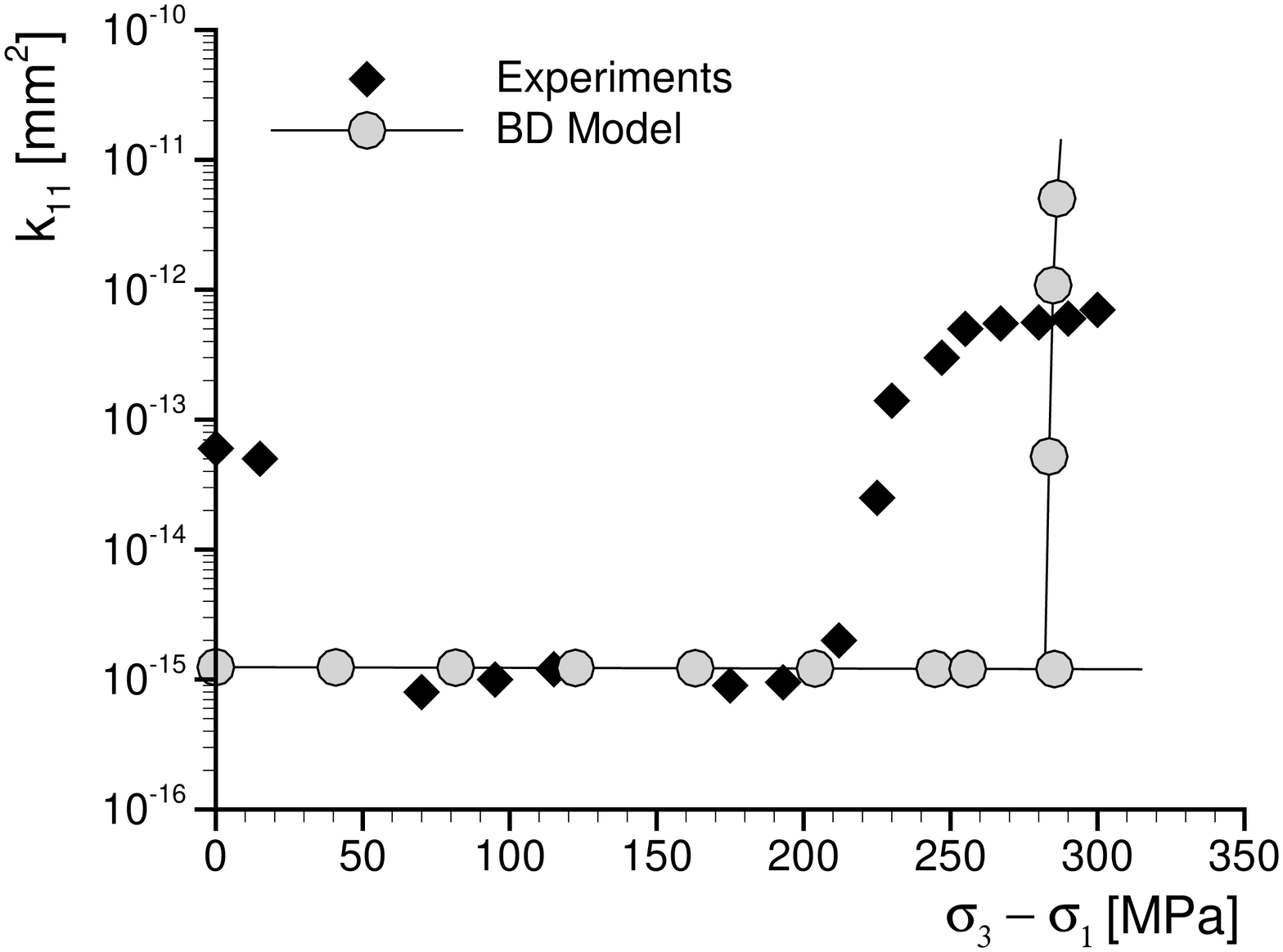,width=0.43\textwidth}\label{Fig:LacDuBonnetb}}
    \subfigure[Beishan granite stress-strain]
    {\epsfig{file=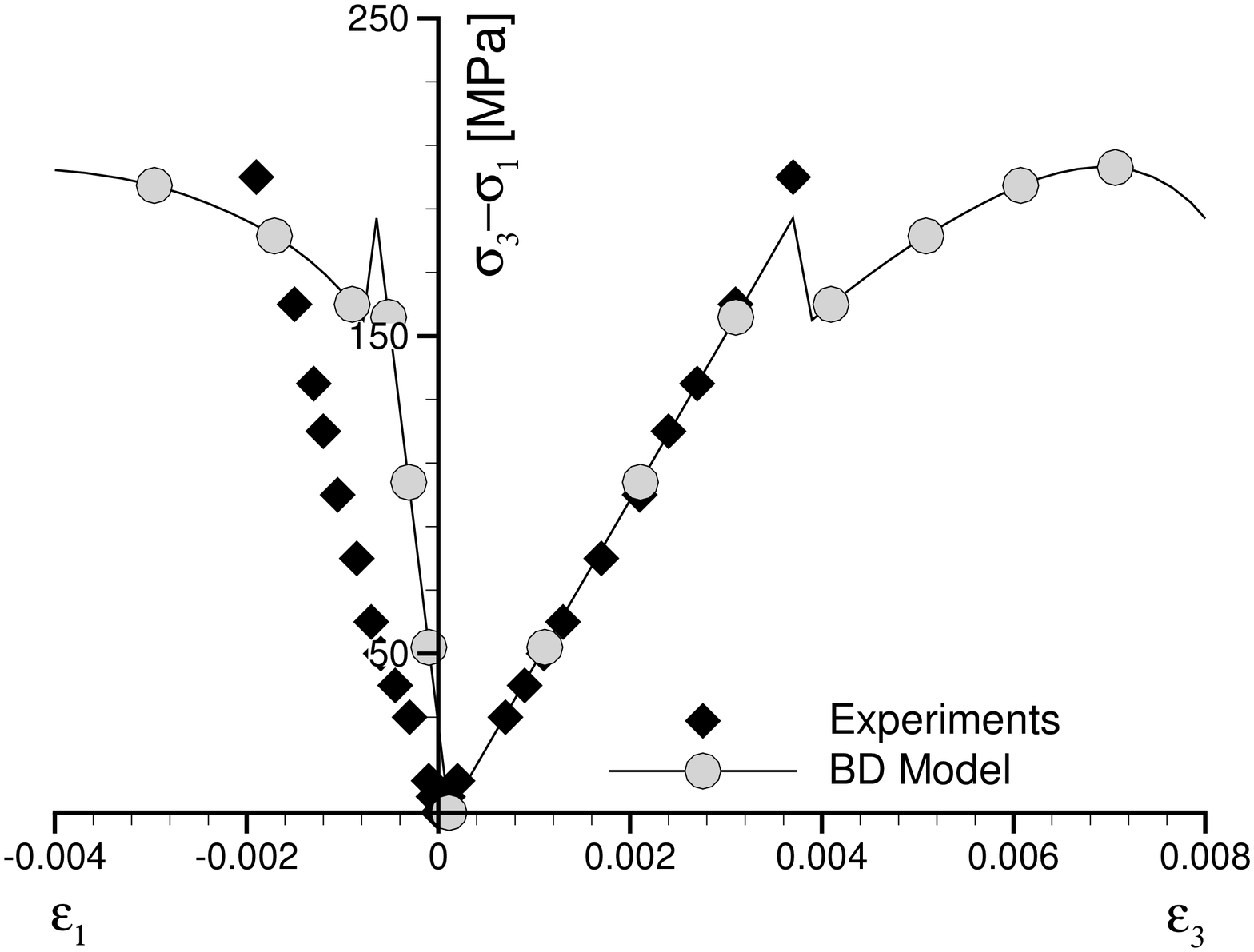,width=0.48\textwidth}\label{Fig:Beishana}}
    \subfigure[Beishan granite permeability]
    {\epsfig{file=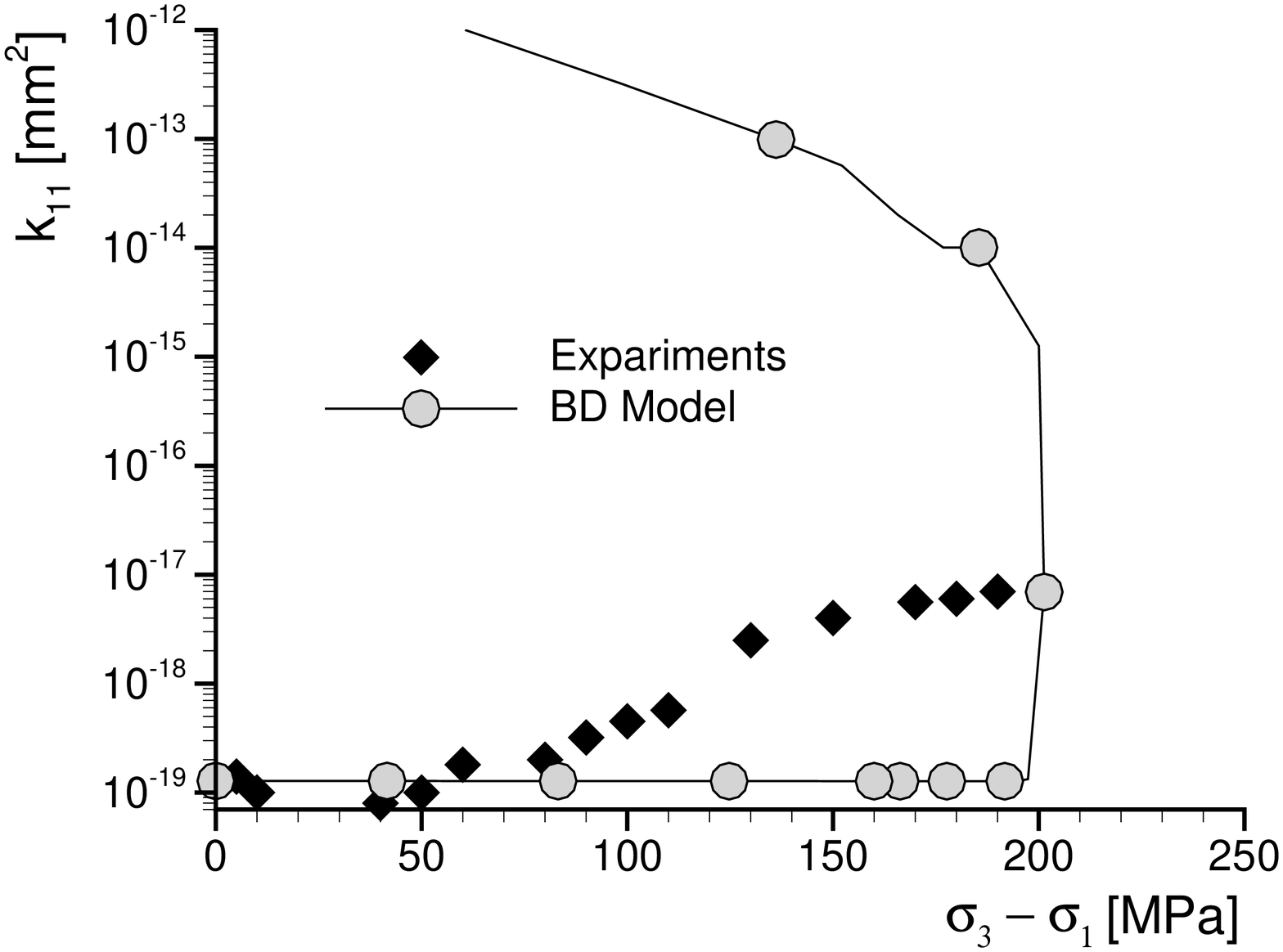,width=0.43\textwidth}\label{Fig:Beishanb}}
    \caption{\small Validation of the brittle damage model versus the experimental tests on Lac du Bonnet \cite{Souley_2001} and Beisahn \cite{Ma_2012} granites . The sample is confined with a 10 MPa pressure. (a)-(c) Deviatoric stress - strain behavior. (b)-(d) Permeability variation with the deviatoric stress.}
    \label{Fig:Granites}
\end{center}
\end{figure}

Experimental data of triaxial compression tests with permeability measurements are available form the literature. In this section we validate the porous brittle damage model, considered at the constitutive level, against experimental results of tests performed on different materials. To facilitate the comparison with the experiments and according to the typical conventions of geomechanics, in this section we assume compressive strains and stresses as positive. Material constants used in the validation are listed in Table~\ref{Table:RockProperties}. Elastic modulus $E$, Poisson coefficient $\nu$, friction angel $\varphi$, porosity $n_0$ and permeability $k_0$ of the intact rocks were recovered from the experimental papers. From $E$ and $\nu$ we computed the Lam\'e constants needed by the elasticity model through the relations
\begin{equation*}
    \lambda = \frac{\nu E}{(1 + \nu)(1 - 2\nu)},
    \qquad
    G = \frac{E}{2(1 + \nu)} \, .
\end{equation*}
The cohesive strength $T_c$ has been derived from the friction angle and the cohesion of the material, while the critical energy release rate $G_c$, not available from the experimental papers, has been calibrated through preliminary analyses. The Kozeny-Carman coefficient has been set zero.
\begin{table}
\begin{center}
\caption {Material constants used for the validation of the material model}
\vskip 0.1in
\begin{tabular}{ccccccccccc}
    \hline
    Rock & $E$ (MPa) & $\nu$ & $T_c$ (MPa) & $G_c$ (N/mm) & $\varphi$ ($^\circ$) & $k_0$ (mm$^2$) & $n_0$ & $C_{\rm KC}$ \\ \hline
    Lac du Bonnet granite \cite{Souley_2001} & 68,000 & 0.21 & 50 & 10 & 46.4 & 10$^{-13}$ & 0.20 & 0.2 \\
    Beisahn granite \cite{Ma_2012}           & 52,000 & 0.21 & 60 & 10 & 35.0 & 10$^{-20}$ & 0.08 & 0.2 \\
    Berea sandstone  \cite{Zhu:1997}         &  8,000 & 0.18 & 50 & 50 & 29.0 & 10$^{-5}$ & 0.21 & 0.2 \\ \hline
\end{tabular}
\label{Table:RockProperties}
\end{center}
\end{table}

We begin with the simulation of the triaxial tests on samples of Lac di Bonnet and Beishan granites documented in \cite{Souley_2001, Ma_2012}. The tests consisted of the application of a confining pressure of 10~MPa, followed by an axial compressive load up to failure. Experiments included the measurement of the permeability of the samples, limited to the pre-peak phase. We simulate the triaxial test with the brittle damage model and compare our numerical results with experiments. Fig.~\ref{Fig:Granites} shows the deviatoric stress, $\sigma_3 - \sigma_1$ versus axial and lateral deformations, $\varepsilon_3$ and $\varepsilon_1$, respectively, and the permeability versus deviatoric stress. During the simulated axial compression, both granites develop one family of faults in shear. The failure plane of the faults corresponds to the one predicted by the Mohr-Coulomb criterion, inclined of an angle $\pi/4 - \varphi/2 $ with respect to direction of maximum stress (21.8$^\circ$ for Lac du Bonnet and $27.5^\circ$ for Beishan). The peak of resistance corresponds to the experimental values, but the brittle damage model predicts a post-peak behavior which is not available in the experimental papers. Experiments show an initial reduction of the permeability, due to the compression of the matrix, followed by a marked increase when the samples begin to show a reduction of stiffness. By contrast, the brittle damage model predicts a constant permeability, which does not increase even after the formation of the shear faults. However, when the load becomes too high to be balanced by friction and the axial loading reduces, faults open and the permeability increases, showing a characteristic behavior often reported in experimental literature, cf. \cite{heiland:2003} and the numerous references therein.
The model is able to capture the dependence of the permeability on porosity and on deformation mechanisms, observed typically in low porosity rocks, where additionally dilatancy is observed when rock fails by brittle faulting \cite{Zhu:1997}. Indeed, microstructural observations have clarified that dilation of the pore volume is primarily due to stress-induced microcracking, which increases permeability by widening the apertures and enhancing the connectivity of the flow paths.

\begin{figure}[!ht]
\begin{center}
    \subfigure[Stress-Strain]
    {\epsfig{file=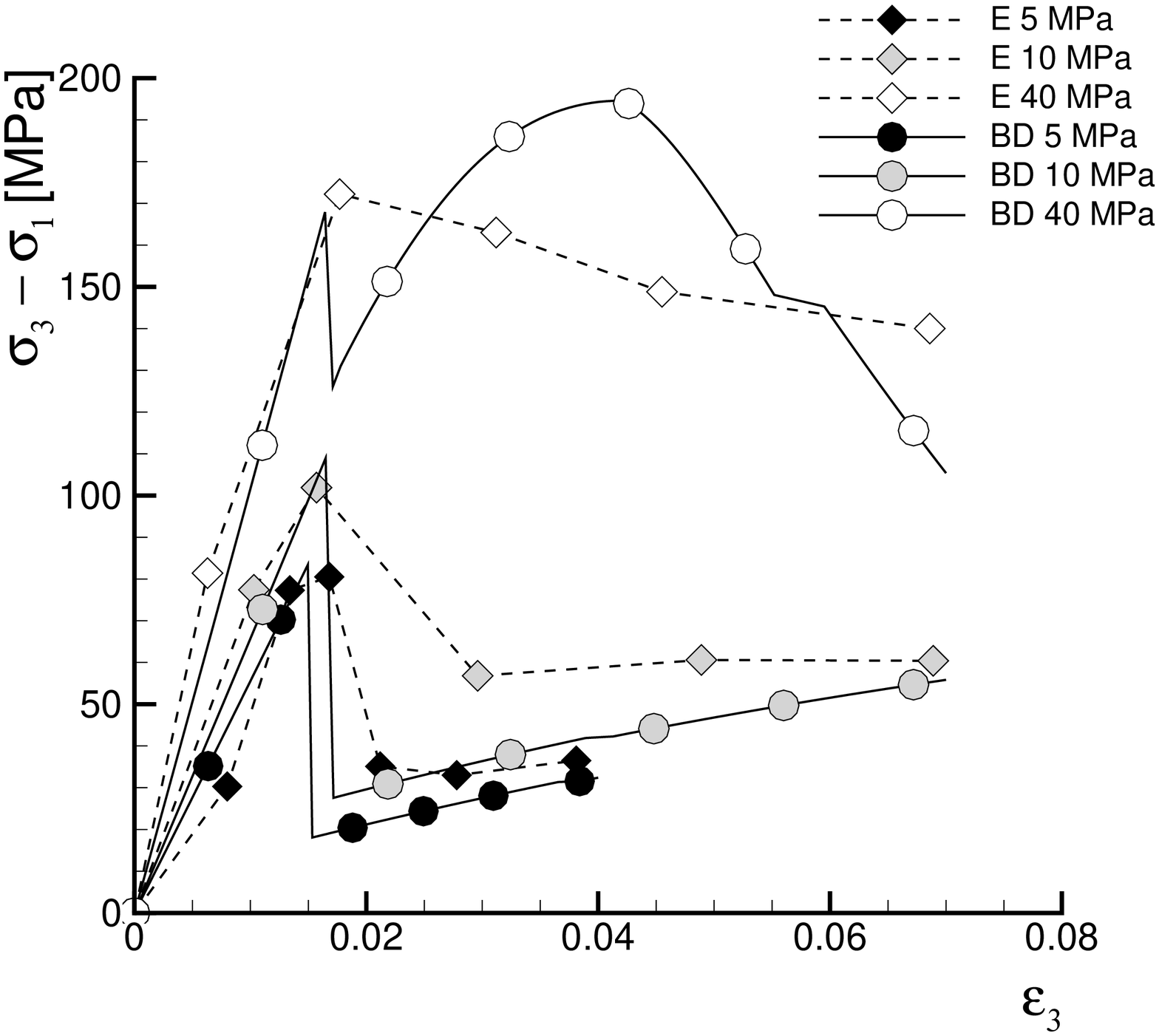,width=0.44\textwidth}\label{Fig:bereaStress}}
    \subfigure[Permeability]
    {\epsfig{file=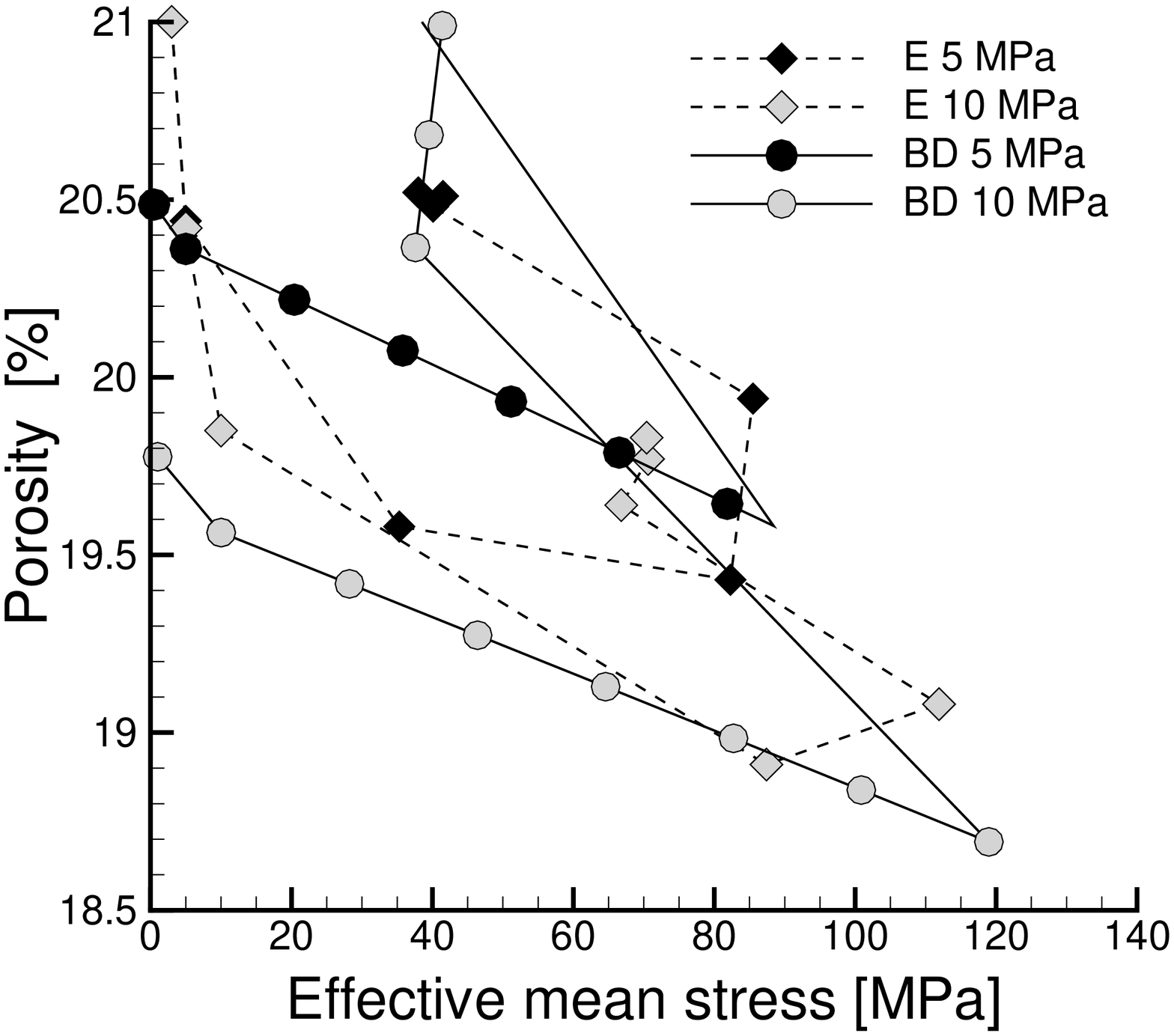,width=0.44\textwidth}\label{Fig:bereaPoro}}
    \subfigure[Stress-Strain]
    {\epsfig{file=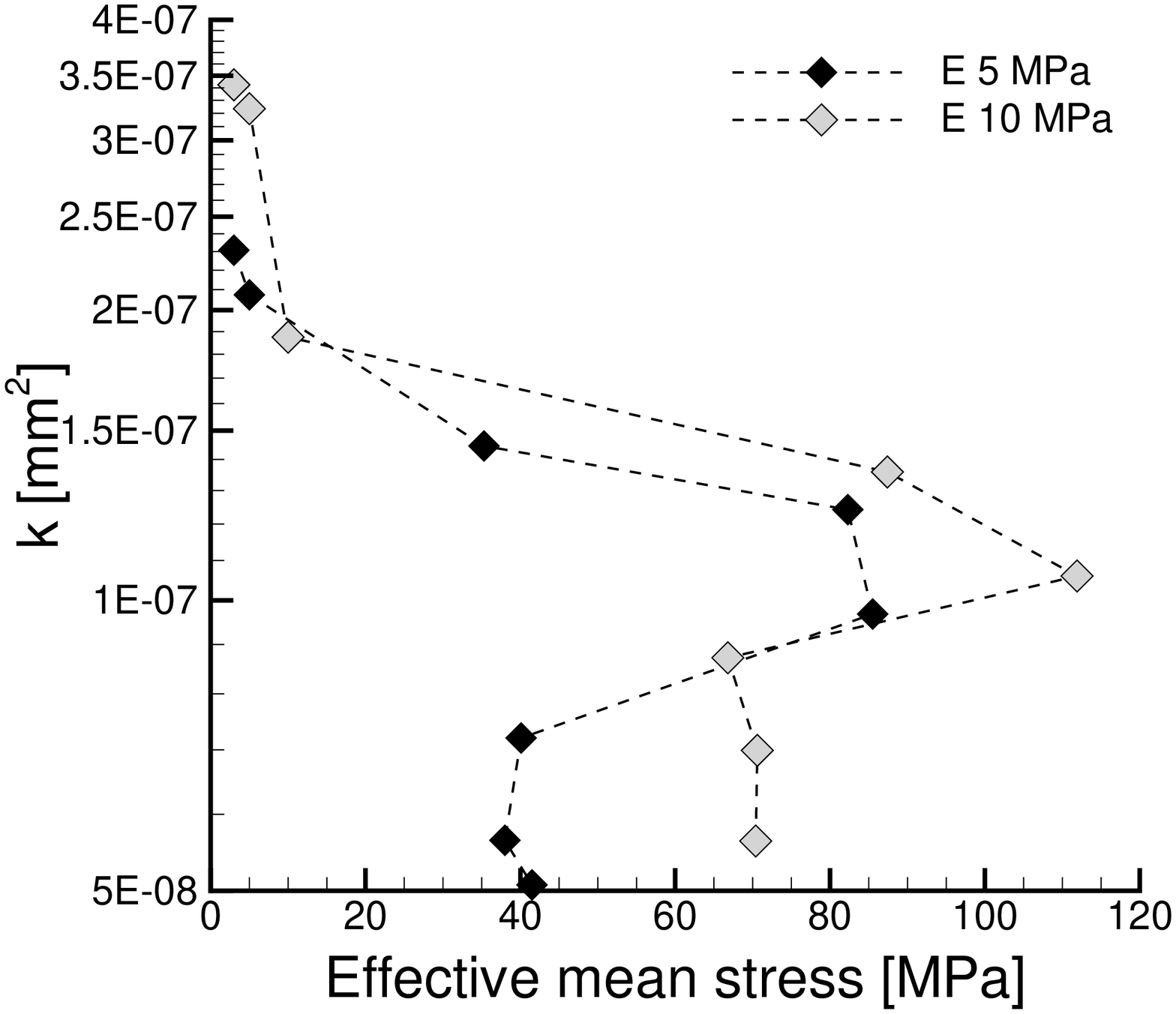,width=0.44\textwidth}\label{Fig:bereaPermExp}}
    \subfigure[Permeability]
    {\epsfig{file=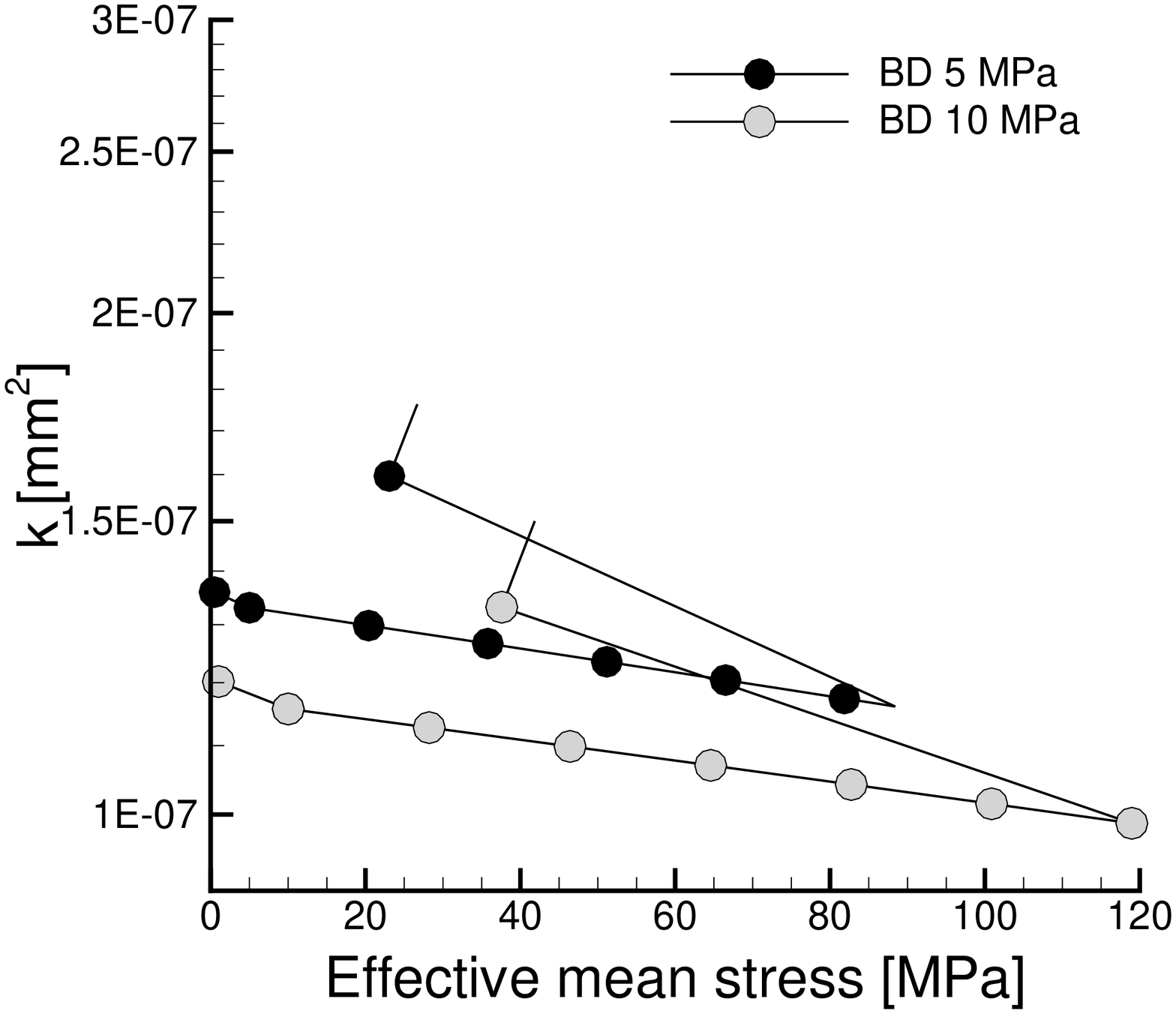,width=0.44\textwidth}\label{Fig:bereaPermNum}}
    \caption{\small Validation of the brittle damage model versus the experimental tests on Berea sandstone \cite{Zhu:1997}. Sample are confined with a 5, 10 or 40 MPa pressure. (a) Deviatoric stress-axial strain behavior. (b) Porosity variation with the deviatoric stress. (c) Experimental permeability, variation with the deviatoric stress. (d) Numerical Permeability, variation with the deviatoric stress.}
    \label{Fig:Sandstone}
\end{center}
\end{figure}
In high porosity rocks, such as sandstones, the effect of stress on permeability is still far to be fully clarified. Literature data testify apparently contradictory observations in the brittle regime \cite{Zhu:1997}. Among the triaxial experiments on Berea sandstone with different confinement reported in \cite{Zhu:1997}, we selected three small confinement triaxial experiments, characterized by a softening stress-strain curve. Pre and post-peak porosity and permeability data are included in the experimental paper. We simulated the experimental tests at confining pressures of 5, 10 and 40~MPa. Experimental and numerical results are shown in Fig.~\ref{Fig:Sandstone}. Fig.~\ref{Fig:bereaStress} shows the deviatoric stress versus the axial deformation. Simulations capture nicely the peak stress for the three tests, while the softening branch is not perfectly reproduced. Fig.~\ref{Fig:bereaPoro} compares numerical and experimental porosity for the two tests at lower confinement pressure. In both simulation and experiment, porosity reduces progressively until the stress peak is reached, and grows during the softening phase, in correspondence to the reduction of the deviatoric stress. Simulations predict qualitatively and quantitatively the variation of porosity during the test. Contrariwise, the comparison between model predictions and experimental observations in terms of permeability evolution is not satisfactory, even from the qualitative point of view. In the experiments, permeability decreased markedly after the stress peak, showing a marked negative correlation between permeability and porosity changes Fig.~\ref{Fig:bereaPermExp}. Experimental data on different sandstones are qualitatively similar, suggesting that permeability evolution as a function of porosity does not follow any systematic trend. A possible explanation of the observed permeability reduction in particular dilating sandstone is that microcracking dramatically increases the tortuosity of the pore space \cite{Zhu:1997}. The brittle damage model predicts a post-peak increase in permeability, see Fig.~\ref{Fig:bereaPermNum}, which is opposite to the Berea sandstone experiments, but in line with many experimental results on low permeability geomaterials, and is also in agreement with the simulations on granites discussed here.

\section{Conclusions}
\label{sec:conclusions}

We have developed a model of distributed fracturing of rock masses, and the attendant permeability enhancement thereof, based on an explicit micromechanical construction resulting in complex connected patterns of cracks, or faults. The approach extends the multi-scale brittle damage material model introduced in \cite{pandolfi:2006}, which was limited to mechanical damage. The fracture patterns that form the basis of the theory are not implied but explicitly defined and the rock mass undergoes throughout compatible deformations and remains in static equilibrium, not just on average at the macroscopic scale, but also the micromechanical level. The sequential faulting construction used to generate the fracture patterns has been shown in \cite{pandolfi:2006} to be optimal as regards the ability of the fracture patterns to relieve stress. In addition, the nucleation criterion, orientation and spacing of the faults derive rigorously from energetic considerations. Following nucleation, fractures can deform by frictional sliding or undergo opening, thereby partially relieving the geostatic stresses in the rock mass. The extension of the theory presented in this paper additionally accounts for fluid pressure by recourse to Terzaghi's effective stress principle. Specifically, we estimate the permeability enhancement resulting from fracture enhancement using standard lubrication theory \cite{snow:1965,snow:1969,parsons:1966}. This extension gives rise to a fully-coupled hydro-mechanical model.

The formulation has been derived in finite kinematics to be consistent with the formulation of the damage model in [18]. A finite kinematics approach is able to describe both large and small strains, so that the model can be applied also to porous media different from rocks. A linear version of the model is currently under development, in view of heavy numerical applications in field problems.

The dry mechanical aspects of the model were validated in \cite{pandolfi:2006} by means of comparisons with the dynamic multiaxial compression experiments on sintered aluminum nitride (AlN) of Chen and Ravichandran \cite{chen:1994, chen:1996, chen:1996b, chen:2000}. The model was shown to correctly predict the general trends regarding the experimental observed damage patterns, as well as the brittle-to-ductile transition resulting under increasing confinement. The hydro-mechanical coupled model has been validated against three different sets of experimental data concerned with triaxial tests at different confinement pressure on granite and sandstone, including Lac du Bonnet \cite{Souley_2001} and Beisahn \cite{Ma_2012} granites and Berea sandstone \cite{Zhu:1997}. The ability of the model to reproduce qualitatively the experimental peak strength, post-peak stress-strain behavior, and permeability enhancement during loading and recovery during unloading is remarkable.

The present coupled hydro-mechanical model has potential for use in applications, such as rocks under geostatic conditions, gravity dams, hydraulic fracture operations, and others, in which a solid deforms and undergoes extensive fracture under all-around confinement while simultaneously being infiltrated by a fluid. The particular case of hydraulic fracture is characterized by the injection of fluid at high pressure, which actively promotes the fracture process and the transport of fluid into the rock mass. Under such conditions, the present model is expected to predict the development of three-dimensional fracture patterns of great complexity over multiple scales. Such complex fracture patterns have indeed been inferred from acoustic measurements in actual hydraulic fracture operations \cite{Warpinski:2005, Wu:2012} and are in sharp contrast to traditional models of hydraulic fracture, which posit the formation of a single mathematically-sharp crack. The present model thus represents a paradigm shift from said traditional models in its ability to account for complexity in the fracture pattern over multiple scale while simultaneously supplying macroscopic effective properties such as permeability and strength that can in turn be used, e.~g., in full-field finite element simulations.

\section*{Appendix A}
\label{AppendixA}

\noindent By denoting the time derivative with $\dot n$, we write
\begin{equation*}\label{app:porosity}
    V \, n = V_V \, ,
    \qquad
    \dot V \, n + V \dot n = \dot V_V \,
    \qquad
    \dot n = \frac{\dot V_V}{V} - \frac{\dot V}{V} \, n \, .
\end{equation*}
Assuming that the solid volume variation is small with respect to the void volume, $\dot V \approx \dot V_V$ and $\dot J \approx \dot J_V = \dot V_V/V_0$. Thus the rate of porosity change becomes:
\begin{equation*}\label{app:porosityRate3}
    \dot n
    =
    \left(1 - n\right) \frac{\dot V_V}{V_0} \frac{V_0}{V}
    =
    \left(1 - n\right) \frac{\dot J_V}{J_V} \, .
\end{equation*}
This relation can be alternatively written in the form
\begin{equation*}\label{app:porosityIntegration}
    \frac{\dot n}{1 - n}
    =
    \frac{\dot J}{J},
    \qquad
    - \log ( 1 - n) = \log J + C \, ,
\end{equation*}
where $C$ is a constant, which can be derived by setting as initial values $J_0 = 1$ and $n_0$, obtaining
\begin{equation*}\label{app:porosityJacobian}
    n = 1 - \frac{1}{J}(1 - n_0) \, .
\end{equation*}

\end{document}